\documentclass[aps,pra,twocolumn,showpacs,eqsecnum,floatfix]{revtex4-1}

\usepackage{amsmath}
\usepackage{amssymb}
\usepackage{graphicx}
\usepackage{epstopdf}
\usepackage{color}
\usepackage[colorlinks, linkcolor=blue, citecolor=blue, urlcolor=blue,breaklinks=true]{hyperref}

\newcommand{\cl}[1]{\mathcal{#1}} 
\newcommand{\on}[1]{#1^\dag \, #1} 
\newcommand{\comm}[2]{\left[\, #1 \, , \, #2 \, \right]} 
\newcommand{\ev}[1]{ \left \langle #1 \right \rangle } 

\begin{document}

\title{Damping of quasiparticles in a Bose-Einstein condensate coupled to an optical cavity}
\author{G. K\'onya}
\affiliation{Institute for Solid State Physics and Optics, Wigner Research Centre, Hungarian Academy of Sciences, H-1525 Budapest P.O. Box 49, Hungary}
\author{G. Szirmai}
\affiliation{Institute for Solid State Physics and Optics, Wigner Research Centre, Hungarian Academy of Sciences, H-1525 Budapest P.O. Box 49, Hungary}
\author{P. Domokos}
\affiliation{Institute for Solid State Physics and Optics, Wigner Research Centre, Hungarian Academy of Sciences, H-1525 Budapest P.O. Box 49, Hungary}

\begin{abstract}
We present a general theory for calculating the damping rate of elementary density wave excitations in a Bose-Einstein condensate strongly coupled to a single radiation field mode of an optical cavity.  Thereby we give a detailed derivation of the huge resonant enhancement in the Beliaev damping of a density wave mode, predicted recently by K\'onya et al., Phys.~Rev.~A 89, 051601(R) (2014). The given density-wave mode constitutes the polariton-like soft mode of the self-organization phase transition. The resonant enhancement takes place, both in the normal and ordered phases, outside the critical region. We show that the large damping rate is accompanied by a significant frequency shift of this polariton mode. Going beyond the Born-Markov approximation and determining the poles of the retarded Green's function of the polariton, we reveal a strong coupling between the polariton and a collective mode in the  phonon bath formed by the other density wave modes.

\end{abstract}


\maketitle

\section{Introduction}

Well-established properties of  ultracold atoms are drastically altered when the atoms are coupled to the radiation field of an optical resonator \cite{ritsch2013cold}. Even if the absorption is suppressed by using only far detuned laser sources, the ensemble of atoms can represent a significant optical density which leads to a strong effect on the field of a high-finesse resonator. The back-action of the cavity field onto the atom cloud is the origin of various novel features or even phenomena. For example, the optical dipole potential exerted dynamically by the cavity field can vary considerably over the kinetic energy scale of the ultracold gas. In this limit, the phase diagram of  strongly localized particles is greatly enriched with respect to the one obtained from the Bose-Hubbard model for an inert external potential \cite{Maschler2008Ultracold,Larson2008Quantum,Vukics2009Cavity,Vidal2010Quantum,li2013lattice}.  In the opposite limit, {\it i.e.}, when the optical dipole potential is negligible and the ultracold atoms form a Bose-Einstein condensate (BEC) which is homogeneous on the optical wavelength scale, the cavity field can still give rise to a significant effect on the elementary excitations, or as often termed ``quasiparticles''. Quasiparticle features are of central importance in general for the description of dynamical many-body phenomena.  A prominent example is the critical mode softening which accompanies the  recently observed  self-organization phase transition  \cite{nagy2010dicke,baumann2010dicke}. 

The system of BEC in an optical resonator proved to be suitable for the quantum simulation of the Dicke model by representing the spin of the original formulation by two collective motional modes of the cloud  \cite{nagy2010dicke,baumann2010dicke}. The Dicke model predicts a critical point when the coupling strength reaches the geometric mean of the frequencies characteristic to the spin and to the boson mode \cite{Emary2003Chaos}. This quantum criticality is the zero temperature limit of the spatial self-organization phase transition of atoms in a cavity \cite{Domokos2002Collective} that has been observed in experiments \cite{Black2003Observation,Arnold2012Selforganization}. Quantum criticality has been observed also in other closely related experiments \cite{Schmidt2014Dynamical,Kessler2014Optomechanical} where one can invoke  a variant of the Dicke model as a few-mode, simplified model to interpret the observations. There are also many theoretical generalizations to describe other exotic phases \cite{Baksic2014Controlling}, such as magnetism \cite{safaei2013raman}, glassiness \cite{gopalakrishnan2009emergent,Strack2011Dicke,Jing2011Quantum,buchhold2013dicke,Habibian2013Boseglass}, or related self-ordering criticality with fermionic atoms \cite{piazza2014umklapp,Keeling2014Fermionic,Chen2014Superradiance}.

Critical behaviour in quantum phase transitions is determined by the dynamical features of the soft mode. In an open system the set of relevant parameters is expanded by the properties of the  driving and dissipation channels. The system of  laser-illuminated atoms coupled to a cavity mode realize, in fact, an open system variant of the Dicke model \cite{Dimer2007Proposed,Baden2014arXiv,buchhold2013dicke}. Indeed, as it has been predicted \cite{nagy2011critical,oztop2012excitations} and recent experiments have shown \cite{Brennecke2013Realtime},  dissipation and the accompanying quantum fluctuations substantially modify the correlation functions and the critical exponents \cite{DallaTorre2010Quantum,bastidas2012nonequilibrium,dalla2013keldysh}. Dissipation is thus a key player in quantum phase transitions \cite{Morrison2008Dynamical,Kessler2012Dissipative,Grimsmo2013Dissipative,Keeling2010Collective,Bhaseen2012Dynamics,Kopylov2013Counting,piazza2013bose,eleuch2014open,xu2013simulating}. 

The experiment performed by Brennecke et al.~\cite{Brennecke2013Realtime} revealed that the interaction between the quasiparticles in a BEC  is relevant to quantitatively interpret measurement data on the superradiant phase transition of the Dicke-model.  Motivated by this observation we generalized the previous models so that to include other dissipation channels that can play a non-negligible role.
In the special case under consideration, the soft mode consists dominantly of a collective density wave excitation of the BEC  \cite{mottl2012roton}.   Therefore, the friction of a density wave quasiparticle in a superfluid of weakly interacting bosonic atoms has to be reconsidered. 

There are basically two collisional mechanisms responsible for the decay of a density wave in a BEC \cite{Ozeri05a}. The first one is Landau damping \cite{Jackson2003Landau, Guilleumas1999Temperature, Guilleumas2003Landau, Tsuchiya2005Landau, Jackson2002Acccidental}, in which the given quasiparticle and another one combine into a third quasiparticle. This mechanism needs a thermal occupation of the other excitation, therefore it vanishes at zero temperature. On the other hand, it exists also  in non-superfluid systems. The second mechanism, characteristic only to superfluids, is Beliaev damping \cite{Hodby2001Experimental, Katz2002Beliaev}. In this case, stimulated by the superfluid background, the selected quasiparticle decays into two lower energy excitations. This process occurs even at zero temperature \cite{Kagan2001Damping}.

In general, the damping rate of quasiparticles that constitute the soft mode is expected to depend on the control parameter of the phase transition. This is simply because the frequency of the soft mode varies over a large range before it vanishes at the critical point.  However, the monotonous variation of the frequency as approaching to the critical point  is accompanied, unexpectedly, by a drastic, resonance-like enhancement in the damping rate \cite{Konya2014Photonic}. Although the mode softening, as we will show, is a necessary ingredient for the effect, the resonant peak is clearly outside the critical region.

In this paper we will present a detailed derivation of this effect that has already been briefly reported in Ref.~\cite{Konya2014Photonic}.  The damping rate enhancement can be attributed to the interaction with the other density wave modes of the condensate  via s-wave collision. These density waves are associated with quasi-momentum modes that form a continuum bath for a large BEC, hence we can evaluate its effect within the Born--Markov approximation.   However, it turns out that the interaction between the soft mode and the other quasiparticles is not so weak and we need to resort to a more accurate analysis which is exempt from the Born approximation underlying the results of Ref.~\cite{Konya2014Photonic}. The presented calculation reveals that the soft mode has a non-negligible influence back on the spectrum of the bath of quasi-momentum modes. That is, the nonlinear s-wave scattering couples significantly other modes into the dynamics, thus the soft mode is one component in a set of interacting bosonic modes.

The rest of the paper is structured as follows. In Sec.~\ref{sec:System}, we will introduce the model for the BEC-cavity system which includes many degrees of freedom of the ultracold atom gas. We will present the equations of motion which allow for describing the system beyond the standard Bogoliubov-type mean field approach. This latter, limited to a linearized treatment of  quantum fluctuations,  is used in Sec.~\ref{sec:Polaritons} to determine the polariton and phonon degrees of freedom which are cross-coupled through the terms higher than first order in quantum fluctuations. The effect of phonons on the polaritons is taken into account by means of a bosonization approximation given in Sec.~\ref{sec:Bosonization}. In Sec.~\ref{sec:LandauBeliaev}, the Beliaev and Landau damping rates are evaluated first within Born--Markov approximation for the phonon bath, and then the Markov approximation is carried out also non-perturbatively by means of the Green's function method. Finally, we summarize the results in Sec.~\ref{sec:Conclusions}. 

\section{Ultracold atoms in an optical resonator}
\label{sec:System}

We consider a Bose-Einstein condensate of ultracold alkali atoms loaded in the volume of a high-finesse, single-mode, optical resonator.  The atoms are illuminated by a far-detuned laser from a direction perpendicular to the cavity axis. The detuning $\Delta_A = \omega - \omega_A$ between the laser and the atomic transition frequency is large enough so that the atoms behave as linear scatterers and their internal dynamics can be adiabatically eliminated. At the same time, the scattering is enhanced in the cavity mode since the driving frequency is close to that of the selected single cavity mode, i.e., the detuning $\Delta_C= \omega - \omega_C$ is on the order of the cavity linewidth  $|\Delta_C| \sim \kappa$. 
 
\begin{figure}[ht]
\begin{center}
\includegraphics[width=\columnwidth]{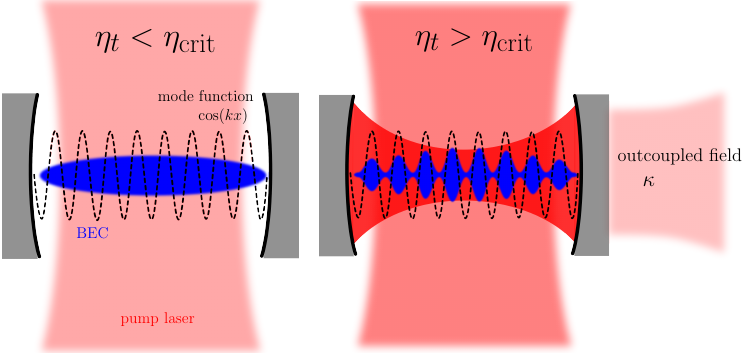}
\end{center}
\caption{(Color online) Schematic representation of the self-organization phase transition. Left: below a threshold value of the transverse laser pump power, the BEC fills the cavity homogeneously on the wavelength scale, and there is no light scattering from the pump laser into the cavity. Above threshold (right panel), the condensate self-organizes into a wavelength periodic pattern and Bragg-scatters into the cavity. The field building up in the cavity traps the atoms in the patterned spatial structure thereby stabilizing the ordered phase.}
\label{experiment}
\end{figure}

Such a transverse pumping geometry is known to exhibit a critical point, as illustrated in Fig.~\ref{experiment}. Below a threshold pump power, a homogeneous Bose-Einstein condensate together with no coherent photons in the cavity remains a stable solution.  This is interesting since the collisional properties and damping of quasiparticles can be studied for the elementary case of a homogeneous superfluid.  When the intensity of the driving laser exceeds a critical value, the condensate density is spatially modulated according to the cavity mode function, and the condensate atoms can coherently scatter photons into the cavity. There appears two stable self-organized solutions connected by a $Z_2$ symmetry, which is spontaneously broken in the high intensity phase. The theory we will develop below applies, of course, also to this inhomogeneous situation. 

The essentials of the self-organization phase transition can be seized by a two-mode approximation, which can be mapped to the Dicke model \cite{nagy2010dicke,baumann2010dicke}. The  measured  phase diagram as well as the spectrum of fluctuations can be interpreted by means of a single  motional mode coupled to the cavity photon mode. Such a simplified approach has been thus verified, although the experiment included effectively a two-dimensional geometry for the cloud. The parameters of the two-mode model, of course, depend on the geometric factors and the dimension of the problem. In the following, we have to resort to a multimode model for describing higher-order than  {usual} mean field effects. However, similarly to the mean-field description of the self-organization phase transition, we will stick to considering only one-dimensional motion of the atoms, which offers the most transparent presentation of the effect of the coupling to photons on the damping properties of superfluid quasiparticles.  {Later, when certain results are of interest also quantitatively, we will consider the question of dimensionality.}

\subsection{Hamiltonian in Bloch-state basis}

The single-mode cavity field is described by the mode function $\cos(k x)$, where $k$ is the wave number, and is associated with the bosonic annihilation and creation operators $a$ and $a^\dagger$. The atomic motion is represented by the second-quantized wavefunction   $\hat \Psi(x)$ and its hermitian conjugate $\hat \Psi^\dagger(x)$. The grand canonical Hamiltonian of the system, in units of $\hbar=1$, in a frame rotating at the laser frequency $\omega$ is given by
\begin{multline} \label{K_coordinate_rep}
\hat{K} = \hat{H}-\mu\hat{N} = -\Delta_C \; \on{\hat{a}} \; + \; \int_0^L \hat{\Psi}^\dag (x) \Biggl[ -\frac{1}{2m}\frac{d^2}{dx^2} - \mu  \\
+ \eta_t \left( \hat{a}^\dag +\hat{a} \right) \cos(k x)
+ U_0 \; \on{\hat{a}} \; \cos^2 (k x) \Biggr] \hat{\Psi}(x) \, dx  \\
+ \; \frac{g}{2} \int_0^L \hat{\Psi}^\dag(x) \hat{\Psi}^\dag(x) \hat{\Psi}(x) \hat{\Psi}(x) \; dx \; .
\end{multline}
The first term is the photon energy in the rotating frame, the detuning $\Delta_C$ must be negative (``red") in order to have a well defined ground state. Next, the spatial integral contains the kinetic energy for particles with mass $m$ and the chemical potential $\mu$. There are three kinds of interaction in the system. The first is connected to the scattering between the laser drive and the cavity mode which is described by the effective amplitude $\eta_t$. The spatial dependence of this interaction inherits the cavity mode function. Note that the time-dependent driving is removed from this term by going to the rotating frame.  The second is the dispersive phase shift exerted by the atoms on the cavity mode resonance, and is characterized by $U_0$ being the resonance shift by a single atom at an antinode. This interaction involves a cavity photon absorption and emission, thus the spatial dependence is $\cos^2(k x)$. Both of these interactions is proportional simply to the matter-wave field density $\hat{\Psi}^\dag (x) \hat{\Psi} (x)$. Finally, the last term is nonlinear in the atom density and accounts for the s-wave collisions between the atoms, the strength is given by $g$. 

The periodicity of the atom-field interaction terms with the wavenumber $k$ suggests that we introduce the Bloch-state basis for the atomic field operator
\begin{multline}
\hat{\Psi}(x) \; = \; \frac{1}{\sqrt{L}} \; \sum_q \; e^{i q x} \Biggl[ \hat{b}_q \, + \\
\, \sqrt{2} \, \cos(k x) \, \hat{c}_q \, + \, \sqrt{2} \, \sin(k x) \, \hat{s}_q \Biggr] \; ,
\end{multline}
where the quasi-momentum is in the interval $q \in \left( -\frac{k}{2} , +\frac{k}{2} \right)$. The lowest band is $b_q$ with homogeneous wavefunction. The first and second excited bands are expanded by combinations of the $c_q$ and $s_q$ modes having $\cos(k x)$ and $\sin(k x)$ wave functions, which are coupled by the kinetic energy term. Modes in these bands carry, beside the quasi-momentum $q$, a momentum $k$ equivalent of the photon wave number. Higher bands are neglected in this study, which is exactly valid below the critical point and is a good approximation  above, but still in the vicinity of the critical point \cite{konya2011multimode}. In brief, the matter-wave field is treated in a three-band approximation \cite{Nagy2013Cavity} instead of the previously used two-mode description \cite{nagy2010dicke,nagy2011critical,konya2012finite}. 

The grand canonical Hamiltonian written in Bloch basis reads as
\begin{equation}
\hat{K}=\hat{K}_{\rm cavity}+\hat{K}_{\rm atoms}+\hat{K}_{\rm pump}+\hat{K}_{\rm disp}+\hat{K}_{\rm coll} \; .
\end{equation}
The cavity Hamiltonian remains the same, 
\begin{equation}
\hat{K}_{\rm cavity} = -\Delta_C \; \on{\hat{a}} \; .
\end{equation}
The atomic Hamiltonian is given by
\begin{multline}
\hat{K}_{\rm atoms} = \sum_q \Biggl[ \left( \frac{q^2}{2m} - \mu \right) \, \on{\hat{b}_q}  \\
 + \left( \frac{k^2+q^2}{2m} - \mu \right) \left( \on{\hat{c}_q} + \on{\hat{s}_q} \right) \\
+ \frac{i q k}{m} \left( \hat{s}_q^\dag \, \hat{c}_q - \hat{c}_q^\dag \, \hat{s}_q \right) \Biggr] \; .
\end{multline}
Note that for $q \neq 0$ the $\hat{c}_q$ and $\hat{s}_q$ modes are coupled.
As a result of scattering a laser photon into the cavity, or reversely,  atoms are transfered between the $\hat{b}_q$ and $\hat{c}_q$ modes 
\begin{equation} \label{K_pump}
\hat{K}_{\rm pump}=\frac{\sqrt{2}}{2} \, \eta_t \, \left( \hat{a}^\dag + \hat{a} \right) \sum_q \left( \hat{b}_q^\dag \, \hat{c}_q + \hat{c}_q^\dag \, \hat{b}_q \right) \; .
\end{equation}
The next dispersive interaction term is proportional to the product of the photon number and the atomic occupation numbers
\begin{equation}
\hat{K}_{\rm disp}=\frac{1}{4} U_0 \, \on{\hat{a}} \, \sum_q \left( 2 \, \on{\hat{b}_q} + 3 \, \on{\hat{c}_q} + \on{\hat{s}_q} \right) \; .
\end{equation}
The collision term consists of two parts,
\begin{equation}
\hat{K}_{\rm coll} = \hat{K}_{\rm normal} + \hat{K}_{\rm umklapp} \; .
\end{equation}
For normal collisions, the quasi-momentum is conserved
\begin{multline}
\hat{K}_{\rm normal}= \frac{g}{2 L} \sum_{q_1 \, q_2 \, q_3 \, q_4} \Biggl[ 
\hat{b}^\dag_{q_1} \, \hat{b}^\dag_{q_2} \, \hat{b}_{q_3} \, \hat{b}_{q_4} \\
+ \frac{3}{2} \, \left( \hat{c}_{q_1}^\dag \, \hat{c}_{q_2}^\dag \, \hat{c}_{q_3} \, \hat{c}_{q_4} 
+ \hat{s}_{q_1}^\dag \, \hat{s}_{q_2}^\dag \, \hat{s}_{q_3} \, \hat{s}_{q_4}  \right) \\
+ \left( \hat{b}_{q_1}^\dag \, \hat{b}_{q_2}^\dag \, \hat{c}_{q_3} \, \hat{c}_{q_4} 
+ \hat{c}_{q_1}^\dag \, \hat{c}_{q_2}^\dag \, \hat{b}_{q_3} \, \hat{b}_{q_4}  \right) \\
+ \left( \hat{b}_{q_1}^\dag \, \hat{b}_{q_2}^\dag \, \hat{s}_{q_3} \, \hat{s}_{q_4} 
+ \hat{s}_{q_1}^\dag \, \hat{s}_{q_2}^\dag \, \hat{b}_{q_3} \, \hat{b}_{q_4}  \right)  \\
+ \frac{1}{2} \, \left( \hat{c}_{q_1}^\dag \, \hat{c}_{q_2}^\dag \, \hat{s}_{q_3} \, \hat{s}_{q_4} 
+ \hat{s}_{q_1}^\dag \, \hat{s}_{q_2}^\dag \, \hat{c}_{q_3} \, \hat{c}_{q_4}  \right)  \\
+ 4 \, \left( \hat{b}_{q_1}^\dag \, \hat{c}_{q_2}^\dag \, \hat{b}_{q_3} \, \hat{c}_{q_4} 
+ \hat{b}_{q_1}^\dag \, \hat{s}_{q_2}^\dag \, \hat{b}_{q_3} \, \hat{s}_{q_4}  \right)  \\
+ 2 \, \hat{c}_{q_1}^\dag \, \hat{s}_{q_2}^\dag \, \hat{c}_{q_3} \, \hat{s}_{q_4} \Biggr] \delta_{\, \tilde{q} \, , \, 0} \; \; \; ,
\end{multline}
where $\; \tilde{q}=q_3+q_4-q_1-q_2 \;$ is the difference between the total incoming and outgoing quasi-momenta. For umklapp processes, the value of the total quasi-momentum changes with $+k$ or $-k$:
\begin{equation}
\hat{K}_{\rm umklapp}= \frac{g}{2 L} \sum_{q_1 \, q_2 \, q_3 \, q_4} \Biggr[ \left( \ldots \right) \delta_{\, \tilde{q} \, , \, k} 
+ \left( \ldots \right) \delta_{\, \tilde{q} \, , \, -k} \Biggr] \; .
\end{equation}
As we will see later, umklapp processes are negligible, so we don't give the detailed expression here.

\subsection{Bose-Einstein condensate in the cavity}

All the system variables can be split to the sum of their expectation values and quantum fluctuations, 
\begin{subequations} \label{expectation_value_fluctuation_decomposition}
\begin{align}
\hat{a} \,   &= \, \sqrt{N_c} \; \alpha \, + \, \tilde{a} \\
\hat{b}_q \, &= \, \sqrt{N_c} \; \beta \, \delta_{q \, , \, 0} \, + \, \tilde{b}_q  \\
\hat{c}_q \, &= \, \sqrt{N_c} \; \gamma \, \delta_{q \, , \, 0} \, + \, \tilde{c}_q \\
\hat{s}_q \, &= \, \tilde{s}_q
\end{align}
\end{subequations}
We assume that the condensate is formed at the center of the lowest band.

The coherent electromagnetic field amplitude in the resonator is $\alpha$. The total number of condensate atoms is $N_c$, which is distributed according to the amplitudes  $\beta$ and $\gamma$ between the homogeneous $b_0$ and the cosine-like $c_0$ modes, respectively. The normalization condition is then $|\beta|^2 + |\gamma|^2 =1$, which allows for determining the chemical potential $\mu$.  The condensate does not extend into the sine-like $s_0$ mode because it is not coupled to the $b_0$ and $c_0$ modes by the coherent atom-photon interactions. This follows simply from the parity conservation of the interaction \eqref{K_coordinate_rep}. The operators denoted by tilde correspond to the fluctuations.

The threshold for the self-organization phase transition is at $\sqrt{2 N_c} \eta_{\rm crit}=\sqrt{-\Delta_C+\frac{1}{2} N_c U_0} \; \sqrt{\omega_R + 2 N_c g /L}$. Below the critical driving, the system is in the normal phase corresponding to the simple solution $\alpha=0$, $\beta=1$, and $\gamma=0$ \cite{nagy2010dicke, baumann2010dicke}. Above threshold, $\gamma$ gradually increases, and far above threshold the approximation of restricting the atomic wavefunction into three bands is no longer valid.

The excitations of the system can be grouped into two sets. For $q=0$, the laser pump couples to the operators $\tilde{a}$, $\tilde{b}_0$ and $\tilde{c}_0$, and these form the \emph{polariton excitations} of the system. The remaining $q \ne 0$ modes, $\tilde{b}_q$, $\tilde{c}_q$ and $\tilde{s}_q$, form  the \emph{phonon excitations}.

It is useful to introduce new parameters for the coupling strengths, 
\begin{subequations}
\begin{align}
y \, &= \, \sqrt{2 \, N_c} \; \eta_t  \\
u \, &= \, \frac{1}{4} \, N_c \, U_0  \\
\tilde{g} \, &= \, \frac{N_c}{L} \, g \; \; ,
\end{align}
\end{subequations}
which have well defined values in the thermodynamic limit, defined as
$N_c \rightarrow \infty, \, L \rightarrow \infty, \, N_c/L= {\rm const}$. Accordingly, the critical coupling is 
\begin{equation}
 \label{eq:y_crit}
y_{\rm crit} = \sqrt{-\Delta_C+ 2 u} \; \sqrt{\omega_R + 2 \tilde{g}}\; ,
\end{equation}
which we will use in the following for scaling the driving strength.

\subsection{Equations of motion beyond the Bogoliubov approximation}

The dynamics of the system is given by the Heisenberg equation of motion:
\begin{equation}
i \frac{d}{dt} \, \hat{O}(t) = \comm{\hat{O}(t)}{\hat{K}} \; .
\end{equation}
 
After we substitute Eq.~(\ref{expectation_value_fluctuation_decomposition}) into this formula, we obtain a hierarchy of terms.
In the standard Bogoliubov approximation, only the zeroth and the first order terms are kept. The mean-field equations are given by the zeroth order terms and the dynamics of the excitations is determined by the first order terms.
Since we aim at describing the polariton-phonon interaction in our model, we have to go one step further and include the second order terms into our description.

The mean-field equations now read
\begin{subequations}
\label{eq:MeanFieldEq}
\begin{multline}
i \frac{d}{dt} \alpha = -\Delta_C \, \alpha + \frac{1}{2} y \left( \beta^* \gamma + \gamma^* \beta \right) \\
 + u \left( 2 |\beta|^2 + 3 |\gamma|^2 \right) \alpha \; ,
 \end{multline}
\begin{multline}
i \frac{d}{dt} \beta = -\mu \, \beta + \frac{1}{2} y \left( \alpha^* +\alpha \right) \gamma + 2u |\alpha|^2 \beta \\
 + \tilde{g} \left( |\beta|^2 \beta + \beta^* \gamma^2 + 2 |\gamma|^2 \beta \right) \; ,
\end{multline}
\begin{multline}
i \frac{d}{dt} \gamma = \left( \frac{k^2}{2m} - \mu \right) \gamma + \frac{1}{2} y \left( \alpha^* + \alpha \right) \beta  \\
+ 3u |\alpha|^2 \gamma + \tilde{g} \left( \frac{3}{2} |\gamma|^2 \gamma + \gamma^* \beta^2 + 2 |\beta|^2 \gamma \right) \; , 
\end{multline}
\end{subequations}
where the back-action of the fluctuations through the expectation value of the second order terms were omitted.
Numerically, we can search for the steady state solution of these equations, where the left hand side is set to zero.

Now, we give the equations of the fluctuations. Let us introduce the compact vector notation for the polariton and phonon variables
\begin{subequations}
\begin{align}
\tilde{v}   &=\left( \tilde{a} \, , \, \tilde{a}^\dag \, , \,  \tilde{b}_0 \, , \, \tilde{b}_0^\dag \, , \, \tilde{c}_0 \, , \, \tilde{c}_0^\dag  \right)^T \\
\tilde{w}(q)&=\left( \tilde{b}_q \, , \, \tilde{b}_{-q}^\dag \, , \, \tilde{c}_q \, , \, \tilde{c}_{-q}^\dag \, , \, \tilde{s}_q \, , \, \tilde{s}_{-q}^\dag \right)^T \,,
\end{align}
\end{subequations}
respectively. The operators in each of these vectors are linearly coupled among each other, and there is a non-linear cross-coupling between the elements of the different vectors
\begin{subequations} \label{v_w_eq_of_motion}
\begin{align}
i \frac{d}{dt} \, \tilde{v}_\mu &= \sum_{\nu} F_{\mu \nu} \, \tilde{v}_{\nu} + \\
+ \frac{1}{\sqrt{N_c}} &\sum_q \sum_{\alpha , \beta} V^{\alpha \beta}_{\mu} \,
\left\{ \tilde{w}^\dag_{\alpha} (q) \tilde{w}_{\beta} (q) - \ev{ \tilde{w}^\dag_{\alpha} (q) \tilde{w}_{\beta} (q) } \right\} \nonumber \\ 
i \frac{d}{dt} \, \tilde{w}_\mu (q) &= \sum_{\nu} G_{\mu \nu} (q) \, \tilde{w}_{\nu} (q)
+ \frac{1}{\sqrt{N_c}} \sum_{\alpha , \beta} \, W^{\alpha \beta}_{\mu} \; \tilde{v}_{\alpha} \, \tilde{w}_{\beta}(q)
\end{align}
\end{subequations}
These equations establish the basis of our calculations in the rest of the paper.  {The linear part, represented by the matrices $F_{\mu \nu}$ and $G_{\mu \nu} (q)$, are treated usually in the Bogoliubov-type mean field descriptions. The additional terms have not yet been investigated in the context of coupled BEC and optical cavity systems.}

Furthermore, we note that there is also a nonlinear polariton-polariton and phonon-phonon interaction in the system, but these effects are neglected in \eqref{v_w_eq_of_motion}. The reason behind this approximation is that (i) the polariton-polariton interaction turns out to be nonresonant, and (ii) the phonon-phonon interaction  does not give a contribution to the polariton damping rate, which we aim to calculate. In fact, the phonon-phonon interaction determines the damping rate of the phonons. Later on, we will introduce this phonon damping as a phenomenological parameter.

\section{Polaritons and phonons}
\label{sec:Polaritons}

In the previous section, we separated the elementary excitations of the system to polariton and phonon sets. There is a linear coupling among the variables within each of these sets in  \eqref{v_w_eq_of_motion}. In the following, we will perform a Bogoliubov-type diagonalization in order to determine the polariton and phonon eigenmodes which are then coupled in higher order interaction terms. 
 
\subsection{Bogoliubov normal modes}

The matrices $F$ and $G(q)$ representing the linear coupling among the polariton-type and the phonon-type modes, respectively, have left and right eigenvectors
\begin{subequations}
\begin{align}
F \; r^{(\mu)} &= \omega_{\mu} \; r^{(\mu)} \\
F^\dag \; l^{(\mu)} &= \omega_{\mu}^* \; l^{(\mu)} \\
G(q) \; c^{(\nu)} (q) &= \omega_{\nu \, q} \; c^{(\nu)} (q) \\
G^\dag (q) \; d^{(\nu)} (q) &= \omega_{\nu \, q}^* \; d^{(\nu)} (q)
\end{align}
\end{subequations}
The polariton and the phonon normal modes are defined then by 
\begin{subequations}
\begin{align}
\tilde{\rho}_{\mu} \; &= \; l^{(\mu)\dag} \cdot \tilde{v} \\
\tilde{\sigma}_{\mu \, q} \; &= \; d^{(\mu)\dag} (q) \cdot \tilde{w} (q) \,,
\end{align}
\end{subequations}
where the $\mu=-3, \; -2, \; -1, \; 1, \; 2, \; 3$ indexes the polariton eigenfrequencies and the phonon bands. As usual for the general Bogoliubov transformation, the normal modes mix the creation and annihilation operators. In order to be able to separately deal with the annihilation and creation processes for polariton and phonon elementary excitations in the following, we make use of the symmetries of the system of equations. 

\begin{figure}[ht]
\begin{center}
\includegraphics[width=0.5\textwidth]{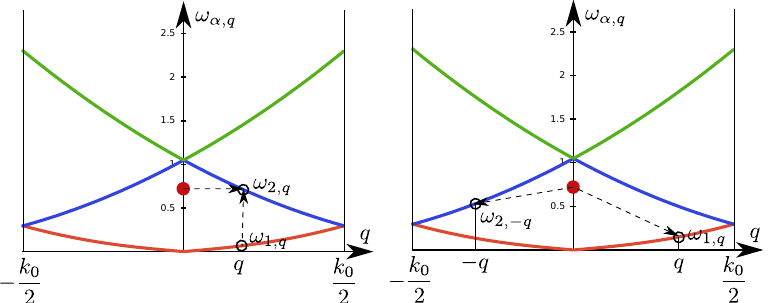}
\end{center}
\caption{(Color online) Phonon spectrum. The laser pump picks the polariton excitation of which frequency is denoted by a red dot. The scattering processes underlying the Landau and Beliaev damping processes are schematically represented on the left and right sides, respectively. }
\label{fig:PhononSpectrum}
\end{figure}

Let us introduce the matrix
\begin{equation}
\Gamma = 
\left( {\begin{array}{*{20}{c}}
   {0} & {1} & {}  & {}  & {}  & {}   \\
   {1} & {0} & {}  & {}  & {}  & {}   \\
   {}  & {}  & {0} & {1} & {}  & {}   \\
   {}  & {}  & {1} & {0} & {}  & {}   \\
   {}  & {}  & {}  & {}  & {0} & {1}  \\
   {}  & {}  & {}  & {}  & {1} & {0}  \\
\end{array}} \right) \; ,
\end{equation}
which simply swaps the creation and the annihilation operators
\begin{subequations}
\begin{align}
\Gamma \cdot \tilde{v}    \; &= \; \tilde{v}^\dag                \\
\Gamma \cdot \tilde{w}(q) \; &= \; \tilde{w}^\dag (-q)   \; ,
\end{align}
\end{subequations}
and where the quasi-momentum is also reflected in the second case. It follows that the matrices $F$ and $G(q)$ have the symmetry, 
\begin{subequations}
\begin{align}
\Gamma \cdot F \cdot \Gamma \; &= \; -  \, F^*        \\
\Gamma \cdot G (q) \cdot \Gamma &= - G^* (-q) \; .  
\end{align}
\end{subequations}
The symmetry $\Gamma$ ensures that the eigenvalues and eigenvectors come in pairs, 
\begin{subequations}
\begin{align}
\omega_{-\mu} \; &= \; -\omega_{\mu}^* \\
r^{(-\mu)} \; &= \; \Gamma \; r^{(\mu)*} \\
l^{(-\mu)} \; &= \; \Gamma \; l^{(\mu)*} \\
\omega_{-\nu \, q} \; &= \; -\omega_{\nu \, -q}^* \\
c^{(-\nu)} (q) \; &= \; \Gamma \; c^{(\nu)*} (-q) \\
d^{(-\nu)} (q) \; &= \; \Gamma \; d^{(\nu)*} (-q) \,.
\end{align}
\end{subequations}
Note that the phonon spectrum is symmetric in the quasi-momentum: $\omega_{\nu \, -q} = \omega_{\nu \, q}$. The phonon spectrum for $\mu=1,2,3$ is plotted in Fig.~\ref{fig:PhononSpectrum}. 

The symmetry guarantees that in a pair of complex eigenvalues the imaginary parts are the same, whereas the real parts have equal magnitude but  opposite sign. We can thus refer to positive and negative frequency modes, according to the sign of the real part of the complex eigenfrequency. For the corresponding eigenvectors, one can prove that $\; \tilde{\rho}_{-\mu}=\tilde{\rho}_{\mu}^\dag \;$ and that $\; \tilde{\sigma}_{-\mu \, q}=\tilde{\sigma}_{\mu \, -q}^\dag \;$. The normal mode expansion can be expressed in terms of only the positive frequency modes,
\begin{subequations}
\label{quasiparticle_expansion}
\begin{align}
\tilde{v} &= \sum_{\mu^+} \left( \tilde{\rho}_{\mu} \cdot r^{(\mu)} \; + \; \tilde{\rho}_{\mu}^\dag \cdot \Gamma \; r^{(\mu)*} \right) \\
\tilde{w} (q) &= \sum_{\mu^+} \left( \tilde{\sigma}_{\mu \, q} \cdot c^{(\mu)} (q) \; + \; \tilde{\sigma}_{\mu \, -q}^\dag \cdot \Gamma \; c^{(\mu)*} (-q) \right) \; \; ,
\end{align}
\end{subequations}
where $\mu^+$ means that we are summing over only the positive frequency modes. The negative modes are automatically included by the second term.  By means of using the $\Gamma$ symmetry, the annihilation and the creation of quasiparticles is manifestly separated in this form.

So far, the symmetry consideration was very general. It relies solely on the fact that the set of variables includes  hermitian conjugate pairs of bosonic annihilation and creation operators, which is then inherited by the Bogoliubov normal modes. To be more specific, here we deal with a Hamiltonian system, which implies an additional symmetry of  the polariton and phonon coupling matrices,  $F$ and $G$, respectively. This symmetry can be formulated by means of the matrix
\begin{equation}
\Omega = 
\left( {\begin{array}{*{20}{c}}
   {+1} & {0} & {}  & {}  & {}  & {}   \\
   {0} & {-1} & {}  & {}  & {}  & {}   \\
   {}  & {}  & {+1} & {0} & {}  & {}   \\
   {}  & {}  & {0} & {-1} & {}  & {}   \\
   {}  & {}  & {}  & {}  & {+1} & {0}  \\
   {}  & {}  & {}  & {}  & {0} & {-1}  \\
\end{array}} \right) \,,
\end{equation}
and reads
\begin{subequations}
\begin{align}
\Omega \cdot F \cdot \Omega \; &= \; F^\dag  \\ 
\Omega \cdot G (q) \cdot \Omega \; &= \; G^\dag (q) \; \; ,
\end{align}
\end{subequations} 
The $\Omega$ symmetry ensures that the eigenfrequencies are real and it also gives a relation between the left and the right eigenvectors:
\begin{subequations}
\begin{align}
\Omega \; r^{(\mu)} &= {\rm sgn}(\omega_{\mu}) \; l^{(\mu)} \\
\Omega \; c^{(\mu)} (q) &= {\rm sgn}(\omega_{\mu \, q}) \; d^{(\mu)} (q) \; \; ,
\end{align}
\end{subequations} 
where ${\rm sgn}( \omega )$ gives the sign of the argument. Since the left and right eigenvectors form a reciprocal basis with respect to each other, we obtain the normalization conditions
\begin{subequations}
\begin{align}
r^{(\mu)\dag} \cdot \Omega \cdot r^{(\nu)} \; &= \; {\rm sgn}(\omega_{\mu}) \; \delta_{\mu \, \nu} \\
l^{(\mu)\dag} \cdot \Omega \cdot l^{(\nu)} \; &= \; {\rm sgn}(\omega_{\mu}) \; \delta_{\mu \, \nu} \\
c^{(\mu)\dag} (q) \cdot \Omega \cdot c^{(\nu)} (q) \; &= \; {\rm sgn}(\omega_{\mu \, q}) \; \delta_{\mu \, \nu} \\
d^{(\mu)\dag} (q) \cdot \Omega \cdot d^{(\nu)} (q) \; &= \; {\rm sgn}(\omega_{\mu \, q}) \; \delta_{\mu \, \nu}
\end{align}
\end{subequations}
With the help of these conditions, one can prove that
\begin{subequations}
\begin{align}
\comm{\tilde{\rho}_{\mu}}{\tilde{\rho}_{\nu}^\dag} \; &= \; \delta_{\mu \, \nu} \;\;\;\;\; \omega_{\mu} \,, \omega_{\nu}>0  \\
\comm{\tilde{\rho}_{\mu}}{\tilde{\rho}_{\nu}} \; &= \; 0  \\
\comm{\tilde{\sigma}_{\mu \, q}}{\tilde{\sigma}_{\nu \, q}^\dag} \; &= \; \delta_{\mu \, \nu} \;\;\;\;\; \omega_{\mu \, q}\,, \omega_{\nu \, q}>0 \\
\comm{\tilde{\sigma}_{\mu \, q}}{\tilde{\sigma}_{\nu \, q}} \; &= \; 0 
\end{align}
\end{subequations}
which verifies that the positive frequency normal modes are bosonic quasiparticles. 

\subsection{Polariton-phonon interaction}

Let us now rewrite the coupled polariton-phonon equations of motion in \eqref{v_w_eq_of_motion} in terms of the 
positive frequency normal modes, i.e., quasiparticles, by using \eqref{quasiparticle_expansion}. 
The equation for the polaritons read ($\omega_\mu > 0$)
\begin{equation}
\label{eq:PolaritonDynamics}
\begin{split}
i \frac{d}{dt} \, \tilde{\rho}_{\mu} \; = \; \omega_{\mu} \, \tilde{\rho}_{\mu} +  \\
 \, + \, \frac{1}{\sqrt{N_c}} \sum_q \sum_{\nu^+ \, \rho^+}
\Biggl[ \; &O^{\mu}_{\nu \, \rho} (q) \cdot \left( \tilde{\sigma}^\dag_{\nu \, q} \; \tilde{\sigma}_{\rho \, q} - \ev{\tilde{\sigma}^\dag_{\nu \, q} \; \tilde{\sigma}_{\rho \, q}} \; \right) \\
 \, + \, \frac{1}{2} \, &M^{\mu}_{\nu \, \rho} (q) \cdot \tilde{\sigma}^\dag_{\nu \, q} \; \tilde{\sigma}^\dag_{\rho \, -q}  \\
 \, + \, \frac{1}{2} \, &N^{\mu}_{\nu \, \rho} (q) \cdot \tilde{\sigma}_{\nu \, -q} \; \tilde{\sigma}_{\rho \, q}
\; \Biggr] \; \; ,
\end{split}
\end{equation}
where the coefficients are given by
\begin{subequations}
\begin{align}
O^{\mu}_{\nu \, \rho} (q) &= \sum_{\alpha \, \beta \, \gamma} \Biggl[ \; l^{(\mu)*}_{\alpha} \cdot V^{\beta \, \gamma}_{\alpha} \cdot c^{(\nu)*}_{\beta \, q} \cdot c^{(\rho)}_{\gamma \, q} \; \nonumber \\ 
+ \; l^{(\mu)*}_\alpha &\cdot V^{\beta \, \gamma}_{\alpha} \cdot \left( \Gamma \; c^{(\rho)} \right)_{\beta \, q} \cdot \left( \Gamma \; c^{(\nu)*} \right)_{\gamma \, q} \; \Biggr] \\
\frac{1}{2} \, M^{\mu}_{\nu \, \rho}  (q) &= \sum_{\alpha \, \beta \, \gamma} \; l^{(\mu)*}_{\alpha} \cdot V^{\beta \, \gamma}_{\alpha} \cdot c^{(\nu)*}_{\beta \, q} \cdot \left( \Gamma \; c^{(\rho)*} \right)_{\gamma \, -q} \\
\frac{1}{2} \, N^{\mu}_{\nu \, \rho}  (q) &= \sum_{\alpha \, \beta \, \gamma} \; l^{(\mu)*}_{\alpha} \cdot V^{\beta \, \gamma}_{\alpha} \cdot \left( \Gamma \; c^{(\nu)} \right)_{\beta \, -q} \cdot c^{(\rho)}_{\gamma \, q}
\end{align}
\end{subequations}
These expressions involve the components of the left- and right eigenvectors of the linear coupling matrices, and the coupling matrix appearing in the original equation \eqref{v_w_eq_of_motion}. All these quantities depend on the mean-field solution, and can be calculated, in general, only numerically. In the first step, the mean field is determined by solving the coupled, nonlinear algebraic equations (\ref{eq:MeanFieldEq}). Then linear matrix algebra is used in a straightforward manner.

Similarly, the phonon equations read ($\omega_\mu > 0$)
\begin{multline}
\label{eq:PhononDynamics}
i \frac{d}{dt} \, \tilde{\sigma}_{\mu \, q} \; = \; \omega_{\mu \, q} \, \tilde{\sigma}_{\mu \, q} +  \\
 \, + \, \frac{1}{\sqrt{N_c}} \sum_{\nu^+ \, \rho^+}
\Biggl[ \; A^{\mu}_{\nu \, \rho} (q)  \cdot \tilde{\rho}_{\nu} \; \tilde{\sigma}_{\rho \, q} 
\, + \, B^{\mu}_{\nu \, \rho} (q) \cdot \tilde{\rho}_{\nu} \; \tilde{\sigma}^\dag_{\rho \, -q} \\
\, + \, C^{\mu}_{\nu \, \rho} (q) \cdot \tilde{\rho}^\dag_{\nu} \; \tilde{\sigma}_{\rho \, q}
\, + \, D^{\mu}_{\nu \, \rho} (q) \cdot \tilde{\rho}^\dag_{\nu} \; \tilde{\sigma}^\dag_{\rho \, -q}
\; \Biggr] \; \; ,
\end{multline}
where the coefficients are
\begin{subequations}
\begin{align}
A^{\mu}_{\nu \, \rho}  (q) &= \sum_{\alpha \, \beta \, \gamma} \; d^{(\mu)*}_{\alpha \, q} \cdot W^{\beta \, \gamma}_{\alpha} \cdot r^{(\nu)}_{\beta} \cdot c^{(\rho)}_{\gamma \, q} \\
B^{\mu}_{\nu \, \rho}  (q) &= \sum_{\alpha \, \beta \, \gamma} \; d^{(\mu)*}_{\alpha \, q} \cdot W^{\beta \, \gamma}_{\alpha} \cdot r^{(\nu)}_{\beta} \cdot \left( \Gamma \; c^{(\rho)*} \right)_{\gamma \, -q} \\
C^{\mu}_{\nu \, \rho}  (q) &= \sum_{\alpha \, \beta \, \gamma} \; d^{(\mu)*}_{\alpha \, q} \cdot W^{\beta \, \gamma}_{\alpha} \cdot \left( \Gamma \; r^{(\nu)*} \right)_{\beta} \cdot c^{(\rho)}_{\gamma \, q} \\
D^{\mu}_{\nu \, \rho}  (q) &= \sum_{\alpha \, \beta \, \gamma} \; d^{(\mu)*}_{\alpha \, q} \cdot W^{\beta \, \gamma}_{\alpha} \cdot \left( \Gamma \; r^{(\nu)*} \right)_{\beta} \cdot \left( \Gamma \; c^{(\rho)*} \right)_{\gamma \, -q} \; \; .
\end{align}
\end{subequations}

We will show in Appendix \ref{V_W_connection_appendix} that the connection between the coefficients $V^{\beta \, \gamma}_{\alpha}$ and $W^{\beta \, \gamma}_{\alpha}$ implies 
\begin{subequations} \label{ABCD_MNO}
\begin{align}
A^{\mu}_{\nu \, \rho}  (q) &= O^{\nu *}_{\rho \, \mu} (q) \\
B^{\mu}_{\nu \, \rho}  (q) &= N^{\nu *}_{\rho \, \mu} (q) \\
C^{\mu}_{\nu \, \rho}  (q) &= O^{\nu }_{\mu \, \rho} (q) \\
D^{\mu}_{\nu \, \rho}  (q) &= M^{\nu }_{\mu \, \rho} (q) \,.
\end{align}
\end{subequations}
This result allows us to introduce an effective Hamiltonian for the polaritons and the phonons, from which the above two equations of motion can be derived as Heisenberg-equations. 

\subsection{Effective Hamiltonian}

The full effective Hamiltonian corresponding to the two equations of motion, Eqs.~(\ref{eq:PolaritonDynamics}) and (\ref{eq:PhononDynamics}), is given by
\begin{equation} \label{full_effective_H}
\begin{split}
\tilde{H} \; = \; \sum_{\mu^+} \omega_{\mu} \, \on{\tilde{\rho}_{\mu}} &+ \sum_q \sum_{\mu^+} \omega_{\mu q} \, \on{\tilde{\sigma}_{\mu \, q}} \\
 \, + \, \frac{1}{\sqrt{N_c}} \sum_q \sum_{\mu^+ \nu^+ \, \rho^+}
\Biggl[ \; &O^{\mu}_{\nu \, \rho} (q) \cdot \tilde{\rho}_\mu^\dag \; \left( \tilde{\sigma}^\dag_{\nu \, q} \; \tilde{\sigma}_{\rho \, q} - \ev{\tilde{\sigma}^\dag_{\nu \, q} \; \tilde{\sigma}_{\rho \, q}} \; \right) \\
 \, + \, \frac{1}{2} \, &M^{\mu}_{\nu \, \rho} (q) \cdot \tilde{\rho}_{\mu}^\dag \; \tilde{\sigma}^\dag_{\nu \, q} \; \tilde{\sigma}^\dag_{\rho \, -q}  \\
 \, + \, \frac{1}{2} \, &N^{\mu}_{\nu \, \rho} (q) \cdot \tilde{\rho}_{\mu}^\dag \; \tilde{\sigma}_{\nu \, -q} \; \tilde{\sigma}_{\rho \, q} + \, h. \, c.
\; \Biggr] \; \; ,
\end{split}
\end{equation}

\begin{figure}[ht]
\begin{center}
\includegraphics[width=0.4\textwidth]{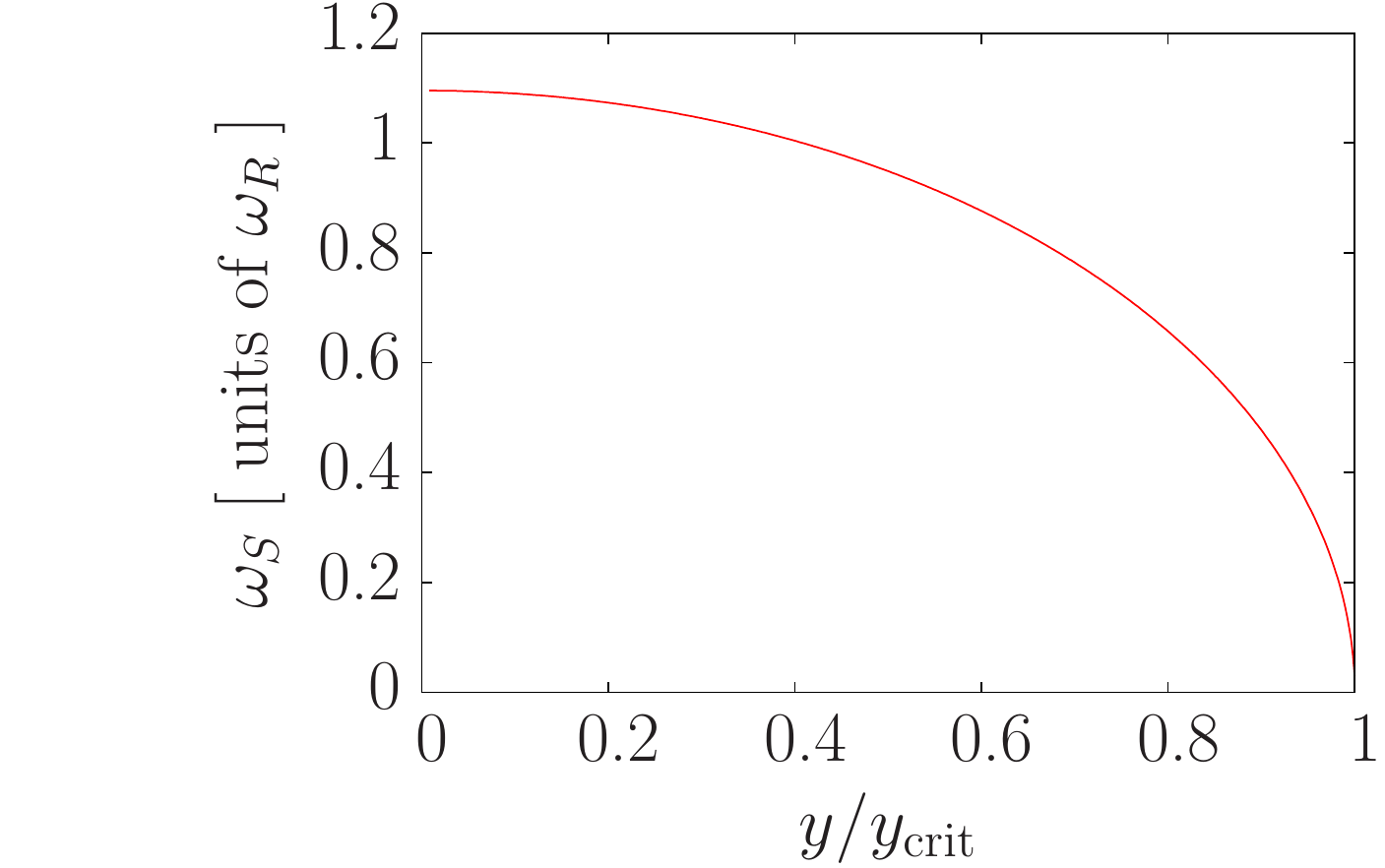}
\end{center}
\caption{(Color online) Real part of the polariton quasiparticle frequency (only the positive frequency part is shown) as a function of the external laser drive strength.  {This polariton is the soft mode of the self-organization phase transition, hence the frequency vanishes at a critical point. Without external driving ($y=0$), the polariton frequency has just the value where the middle and the upper branches touch for $q=0$ in the phonon spectrum in }Fig.~\ref{fig:PhononSpectrum}.}
\label{fig:PolaritonSpectrum}
\end{figure}

So far, we presented a theory which can generally describe the interaction of selected quasiparticles of a cavity-BEC system with the continuum of phonons. In the following we will use  the main results of the theory in an interesting, highly non-trivial case. Now, without losing generality, we will consider only a certain part  of the full effective Hamiltonian, which refers a selected polariton quasiparticle, which is the soft mode of the self-organization phase transition, denoted by $\tilde{\rho}_s$. The frequency of the soft mode as a function of the control parameter normalized to the critical value, $y/y_{\rm crit}$, is plotted in Fig.~\ref{fig:PolaritonSpectrum}). Further, we denote by $\tilde{\sigma}_{1 \, q}$ the lowest, and by $\tilde{\sigma}_{2 \, q}$ the middle phonon branches displayed in Fig.~\ref{fig:PhononSpectrum}.
 
The relevant part of the effective Hamiltonian accounting for the polariton-phonon coupling is
\begin{equation} \label{effective_H_relevant_part}
\begin{split}
\tilde{H} \; &= \; \omega_s \; \on{\tilde{\rho}_s} \; + \; \sum_q \omega_{1 \, q} \; \on{\tilde{\sigma}_{1 \, q}} 
\; + \; \sum_q \omega_{2 \, q} \; \on{\tilde{\sigma}_{2 \, q}} + \\
&+\frac{1}{\sqrt{N_c}} \sum_q \left( g_q^{L} \cdot \tilde{\rho}_s^\dag \, \tilde{\sigma}_{1 \, q}^\dag \, \tilde{\sigma}_{2 \, q} 
\; + \; g_q^{L \, *} \cdot \tilde{\rho}_s \, \tilde{\sigma}_{2 \, q}^\dag \, \tilde{\sigma}_{1 \, q}  \right) \\
&+\frac{1}{\sqrt{N_c}} \sum_q \left( g_q^{B} \cdot \tilde{\rho}_s^\dag \, \tilde{\sigma}_{1 \, q} \, \tilde{\sigma}_{2 \, -q} 
\; + \; g_q^{B \, *} \cdot \tilde{\rho}_s \, \tilde{\sigma}_{2 \, -q}^\dag \, \tilde{\sigma}_{1 \, q}^\dag  \right) \; \; ,
\end{split}
\end{equation}
where $\omega_s$ is the soft mode frequency. The coefficients $g^L_q=O^{s}_{1 2}(q)$ and $g^B_q= \frac{1}{2} N^{s}_{2 1}(q)$ describe the strengths of the so-called Landau- and Beliaev-type coupling processes (illustrated in Fig.~\ref{fig:PhononSpectrum}). In the former, the polariton $\tilde{\rho}_s$ merges with a phonon from the lowest branch to create a phonon on the middle branch. In this process a condensate atom is created simultaneously. The latter, Beliaev process corresponds to the creation of two phonons, this process is stimulated by the background condensate. The energy and the quasi-momentum quantum numbers obviously need to be conserved during these processes. Furthermore the total momentum has to be conserved also, which means that one of the phonons should be in the middle and one should be in the lowest branch.   

The Heisenberg equations of motion generated by this Hamiltonian are nonlinear and cannot be solved generally. In accordance with the usual treatment of open systems and Markov approximation, we will approximate the state of the phonon degrees of freedom as being close to a thermal equilibrium.

\section{Bosonization of the phonon bath}
\label{sec:Bosonization}

Let us introduce two operators which correspond to the Landau and Beliaev processes, respectively,
\begin{subequations} \label{L_B_operators}
\begin{align}
\tilde{L}_q &= \left( \cl{N}_q^{L} \right)^{-1} \cdot \tilde{\sigma}_{1 \, q}^\dag \, \tilde{\sigma}_{2 \, q} \\
\tilde{B}_q &= \left( \cl{N}_q^{B} \right)^{-1} \cdot \tilde{\sigma}_{1 \, q} \, \tilde{\sigma}_{2 \, -q} \; ,
\end{align}
\end{subequations} 
where $\cl{N}_q^{L}$ and $\cl{N}_q^{B}$ are unspecified normalization coefficients.The identity $\comm{A}{BC}=\comm{A}{B}C+B\comm{A}{C}$ implies the algebraic relations 
\begin{subequations}
\begin{align}
\left| \cl{N}_q^{L} \right|^2 \cdot \comm{\tilde{L}_q}{\tilde{L}_q^\dag} &= \on{\tilde{\sigma}_{1 \, q}} - \on{\tilde{\sigma}_{2 \, q}} \\
\left| \cl{N}_q^{B} \right|^2 \cdot \comm{\tilde{B}_q}{\tilde{B}_q^\dag} &= \on{\tilde{\sigma}_{1 \, q}} + \on{\tilde{\sigma}_{2 \, q}} + 1 \\
\cl{N}_q^{B} \, \cl{N}_q^{L} \cdot \comm{\tilde{B}_q}{\tilde{L}_q} &= \tilde{\sigma}_{2 \, -q} \, \tilde{\sigma}_{2 \, q}
\end{align}
\end{subequations}
By assuming that the occupation number in the phonon modes remains close to the thermal one, we can use the following mean field approximation
\begin{subequations}
\begin{align}
\on{\tilde{\sigma}_{\mu \, q}} \; &\simeq \; \bar{n}_{\mu \, q}   \\
\tilde{\sigma}_{2 \, -q} \; \tilde{\sigma}_{2\, q} \; &\simeq \; 0 \; \; , 
\end{align}
\end{subequations} 
where $\bar{n}_{\mu \, q}$ is the thermal occupation number.
By setting the normalization factors as 
\begin{subequations}
\begin{align}
\cl{N}_q^{L} &= \sqrt{ \bar{n}_{1 \, q} - \bar{n}_{2 \, q} \; }   \\
\cl{N}_q^{B} &= \sqrt{ \bar{n}_{1 \, q} + \bar{n}_{2 \, q} + 1 \; } \; \; .
\end{align}
\end{subequations} 
we obtain normal bosonic commutation relations
\begin{subequations}
\begin{align}
\comm{\tilde{L}_q}{\tilde{L}_q^\dag} &= 1 \\
\comm{\tilde{B}_q}{\tilde{B}_q^\dag} &= 1 \\
\comm{\tilde{B}_q}{\tilde{L}_q} &= 0 \; \; .
\end{align}
\end{subequations}
In this approximation scheme, we have introduced new bosonic modes describing the phonons. The effective Hamiltonian can be rewritten as
\begin{equation}
\begin{split}
\tilde{H}_{\rm eff} \; &= \; \omega_s \; \on{\tilde{\rho}_s} \; + \; \sum_q \left( \omega_{2 \, q} - \omega_{1 \, q} \right) \; \on{\tilde{L}_q} \\
&+ \; \sum_q \left( \omega_{1 \, q} + \omega_{2 \, q} \right) \; \on{\tilde{B}_q}  \\
&+\frac{1}{\sqrt{N_c}} \sum_q \cl{N}_q^L \left( g_q^{L} \cdot \tilde{\rho}_s^\dag \, \tilde{L}_q \; + \; g_q^{L \, *} \cdot \tilde{L}_q^\dag \, \tilde{\rho}_s \right) \\
&+\frac{1}{\sqrt{N_c}} \sum_q \cl{N}_q^B \left( g_q^{B} \cdot \tilde{\rho}_s^\dag \, \tilde{B}_q \; + \; g_q^{B \, *} \cdot \tilde{B}_q^\dag \, \tilde{\rho}_s \right) \; \; ,
\end{split}
\end{equation}
where we used the eigenfrequencies of the Landau-type  $\tilde{L}_q$ and Beliaev-type $\tilde{B}_q$ quasiparticles, which come from the definition \eqref{L_B_operators}.  This is now a solvable, quadratic Hamiltonian leading to coupled, linear equations of motion 
\begin{subequations}
\label{eq:LinearizedPolaritonPhononEqs}
\begin{align}
i\frac{d}{dt} \tilde{\rho}_s &= \comm{\tilde{\rho}_s}{\tilde{H}_{\rm eff}} \\
i\frac{d}{dt} \tilde{L}_q &= \comm{\tilde{L}_q}{\tilde{H}_{\rm eff}} - i \left( \gamma_{1 \, q} + \gamma_{2 \, q} \right) \tilde{L}_q + i \, \tilde{\zeta}^{L}_{q} \\
i\frac{d}{dt} \tilde{B}_q &= \comm{\tilde{B}_q}{\tilde{H}_{\rm eff}} - i \left( \gamma_{1 \, q} + \gamma_{2 \, q} \right) \tilde{B}_q + i \, \tilde{\zeta}^{B}_{q} \; \; ,
\end{align}
\end{subequations}
where $\gamma_{\mu \, q}$ is the damping of the phonon mode $\tilde{\sigma}_{\mu \, q}$. Note that we added damping for $\tilde{L}_q$ and $\tilde{B}_q$ together with the accompanying $\tilde{\zeta}_{\, L}$ and $\tilde{\zeta}_{\, B}$ Langevin-type noise terms. The damping rates are the sum of the damping rates of the composite phonon modes. The microscopic calculation of these rates would require a tedious calculation which involves the so-far neglected phonon-phonon coupling terms. Instead of this direct approach, one can use phenomenologically the free-space phonon decay rates, assuming that the phonon decay is hardly affected by the presence of the cavity field.

\section{Landau- and Beliaev-damping}
\label{sec:LandauBeliaev}

The linear set of equations (\ref{eq:LinearizedPolaritonPhononEqs}) can be solved analytically. Since we look for damping rates, or more generally, for the eigenfrequency of the polariton embedded in the phonon bath, we can resort to a Green's function technique. Let us introduce three retarded Green's functions, 
\begin{subequations}
\begin{align}
G^P(t-t')   &= -i \, \theta(t-t') \, \ev{ \comm{\tilde{\rho}_s(t)}{\tilde{\rho}_s^\dag(t')} } \\
G^L_q(t-t') &= -i \, \theta(t-t') \, \ev{ \comm{\tilde{L}_q(t)}{\tilde{\rho}_s^\dag(t')} }  \\
G^B_q(t-t') &= -i \, \theta(t-t') \, \ev{ \comm{\tilde{B}_q(t)}{\tilde{\rho}_s^\dag(t')} } \; \; ,
\end{align}
\end{subequations}
which, after Fourier transformation,
\begin{equation}
f(t-t')= \frac{1}{2\pi} \int d\omega \; f(\omega) \; e^{-i \omega (t-t')} 
\end{equation}
obey a closed set of algebraic equations
\begin{subequations}
\begin{align}
\omega \, G^P(\omega) &= 1 + \omega_s \, G^P(\omega)
+\frac{1}{\sqrt{N_c}} \sum_q g_q^{L} \, \cl{N}_q^L \; G^L_q(\omega) \\
&+\frac{1}{\sqrt{N_c}} \sum_q g_q^{B} \, \cl{N}_q^B \; G^B_q(\omega) \nonumber  \\
\omega \, G^L_q(\omega) &= \omega^L_q \, G^L_q(\omega) + \frac{1}{\sqrt{N_c}} \, g_q^{L \, *} \, \cl{N}_q^L \; G^P(\omega)  \\
\omega \, G^B_q(\omega) &= \omega^B_q \, G^B_q(\omega) + \frac{1}{\sqrt{N_c}} \, g_q^{B \, *} \, \cl{N}_q^B \; G^P(\omega)  \; \; .
\end{align}
\end{subequations}
For brevity, we introduced the complex eigenfrequencies
\begin{subequations}
\begin{align}
\omega^L_q = \left( \omega_{2 \, q} - \omega_{1 \, q} \right) -i \left( \gamma_{1 \, q} + \gamma_{2 \, q} \right) \\
\omega^B_q = \left( \omega_{1 \, q} + \omega_{2 \, q} \right) -i \left( \gamma_{1 \, q} + \gamma_{2 \, q} \right)
\end{align}
\end{subequations}
for the bosons modes $\tilde{L}_q$ and $\tilde{B}_q$, respectively. The polariton Green's function can be expressed in closed form,
\begin{equation}
\label{eq:PolaritonGreen}
G^P(\omega) = \left(\; \omega -\omega_s - \Sigma^L (\omega) -\Sigma^B (\omega) \; \right)^{-1} \; \; ,
\end{equation}
where the two self-energies 
\begin{subequations}
 \label{eq:Selfenergy}
\begin{align}
\Sigma^L(\omega) &= \frac{1}{N_c} \sum_q \left| g_q^{L} \right|^2 \cdot \frac{\left( \cl{N}_q^L \right)^2}{\omega - \omega^L_q} \\
\Sigma^B(\omega) &= \frac{1}{N_c} \sum_q \left| g_q^{B} \right|^2 \cdot \frac{\left( \cl{N}_q^B \right)^2}{\omega - \omega^B_q} \; \; .
\end{align}
\end{subequations}
incorporate the integrated effect of the Landau and Beliaev processes.  {These expressions are the main result of the general theory, in the following we will apply them in special cases relevant to recent experiments. First we will restrict the analysis to the self-energies in order to deduce the damping rate and the frequency shift of the polariton mode within the Born--Markov approximation. Then we will evaluate the polariton Green's function $G^P(\omega)$ which, in principle incorporates the full dynamics of the polariton-phonon system. Finally, we will determine the poles of  complex continuation of the retarded Green's function which reveals the underlying relevant excitations. Since the poles can be far from the real axis, this will turn out to be the case here, one can find significant deviation from the results of the Born--Markov approximation.}

\subsection{Born--Markov approximation}

As a first approximation, the frequency dependence of the self-energies are eliminated ($\Leftrightarrow$ Markov-approximation), and simply its value at the bare system frequency is taken ($\Leftrightarrow$ Born-approximation), 
$\Sigma^L(\omega) \simeq \Sigma^L(\omega_s)$ and $\Sigma^B(\omega) \simeq \Sigma^B(\omega_s)$. The complex eigenfrequency of the polariton mode is identified with the 
pole of the Green's function which is now at
\begin{equation}
 \label{eq:BornMarkovPole}
\omega_{\rm pole}=\omega_s + \Sigma^L(\omega_s) + \Sigma^B(\omega_s) \; \; .
\end{equation}
It follows that the Landau and Beliaev processes give rise to a complex frequency shift
\begin{multline}
 \label{eq:LandauDamping}
\delta^{L} -i \gamma^{L} \; = \; - \frac{1}{N_c} \sum_q \left| g_q^{L} \right|^2  \\
 \times \frac{ \left( \bar{n}_{1 \, q} - \bar{n}_{2 \, q} \right) }{\left( \omega_{2 \, q} - \omega_{1 \, q} -\omega_s \right) -i \left( \gamma_{1 \, q} + \gamma_{2 \, q} \right) } \; \; ,
\end{multline}
and 
\begin{multline}
 \label{eq:BeliaevDamping}
\delta^{B} -i \gamma^{B} \; = \; - \frac{1}{N_c} \sum_q \left| g_q^{B} \right|^2 \\
\times \frac{ \left( \bar{n}_{1 \, q} + \bar{n}_{2 \, q} +1 \right) }{\left( \omega_{1 \, q} + \omega_{2 \, q} -\omega_s \right) -i \left( \gamma_{1 \, q} + \gamma_{2 \, q} \right) } \; \; ,
\end{multline}
respectively. Obviously, the real part corresponds to a frequency shift due to dressing with the phonons, and the imaginary parts correspond to the Landau and Beliaev damping rates.

We evaluate numerically Eqs.~\eqref{eq:LandauDamping} and \eqref{eq:BeliaevDamping}. When performing the quasi-momentum sums, one can use the three-dimensional density of modes instead of the one-dimensional one. To this end, the argument in the summation has to be multiplied by $\frac{1}{2\pi} \left( q w \right)^2$, where $w$ is the width of the condensate. We assign the following numerical values to the parameters $N_c = 10^4$ , $k L/(2\pi) = 1001$, $N_c g/L = 0.1 \omega_R$, $\Delta_C = -1000 \omega_R$ and $k w= 2\pi \, \sqrt{2}$, which corresponds to the experimental values reported in Ref.~\cite{Brennecke2013Realtime}. We will introduce a phenomenological parameter $\epsilon$ for the sum of the damping rates of the two phonons involved in the process, $\epsilon = \gamma_{1 \, q} + \gamma_{2 \, q}$, that is, (i) we neglect the variation of this sum as a function of the quasi-momentum $q$, and (ii)  we renounce to calculate it ab initio from the initial Hamiltonian. In fact, such a calculation would require to keep another second-order phonon-phonon interaction term in Eq.~(\ref{v_w_eq_of_motion}). This term was dropped because it does not give direct contribution to the polariton damping rate. Reversely, the resonator has no considerable effect on the phonon damping (strictly vanishing for a homogeneous condensate below threshold) so that the free-space value could be safely invoked for the calculation.

\begin{figure}[ht]
\begin{center}
\includegraphics[width=\columnwidth]{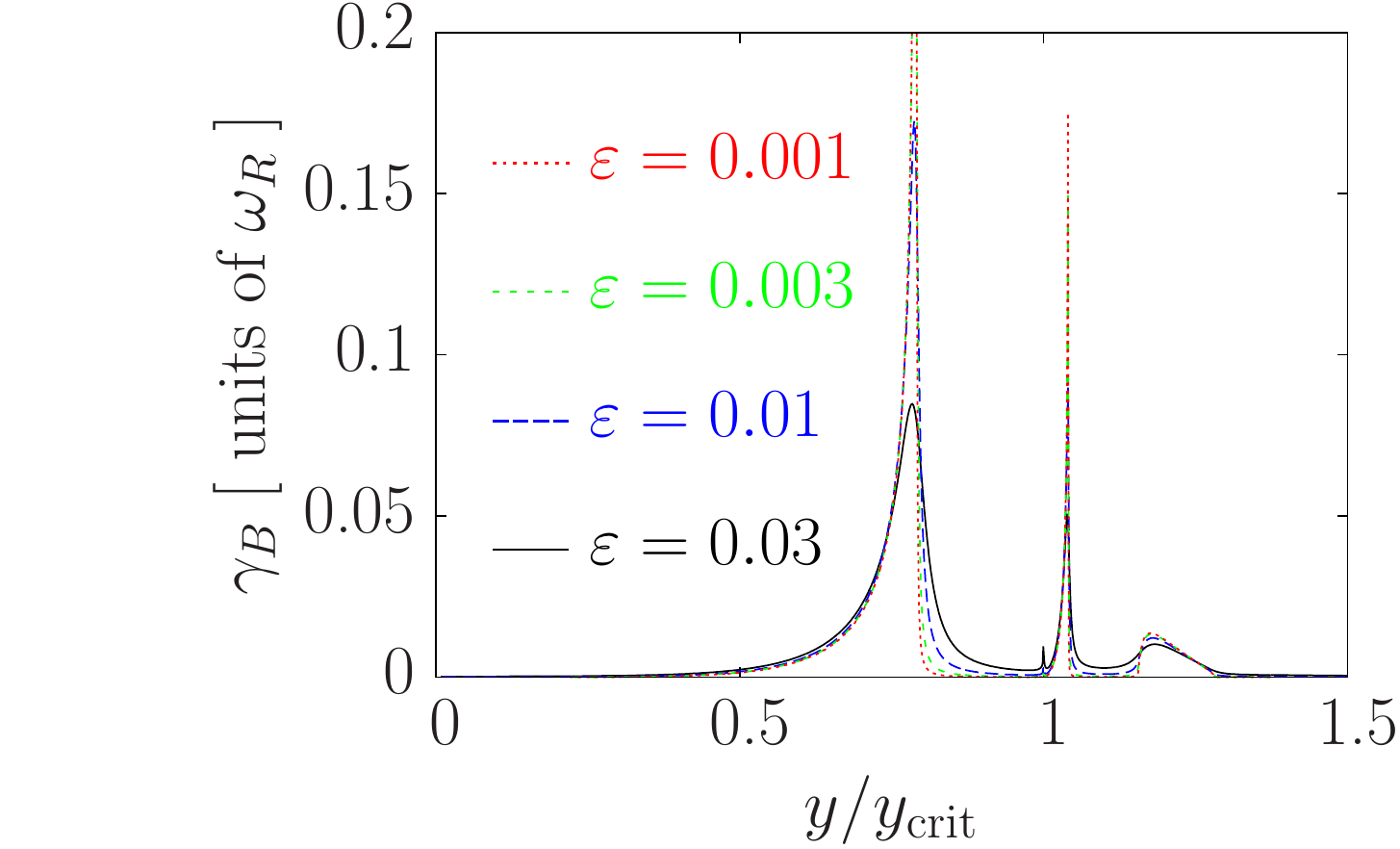}
\end{center}
\caption{(Color online) Beliaev damping rate as a function of the control parameter that is the normalized laser pump strength at zero temperature, $T=0$. There appear resonant peaks both below and above the critical point. For the explanation of their origin, see text below. The quite overlapping curves correspond to various values of the phenomenological phonon damping parameter, $\epsilon= 0.03, 0.01, 0.003, 0.001$ in units of $\omega_R$, in order of increasingly sharp peaks.}
\label{fig:Beliaev_Fugges_kuszob_folott_abra}
\end{figure}
Figure \ref{fig:Beliaev_Fugges_kuszob_folott_abra} shows the damping rate as a function of the control parameter $y$ normalized to the critical value $y_{\rm crit}$. We obtain sharp peaks in the Beliaev damping at certain values of the laser pumping strength. The main reason for the resonant enhancement is connected to the variation of the polariton frequency. The peaks in the damping rate occur when the polariton decays into two phonons being close to the edges of the Brillouin zone $q \approx \pm k/2$. Here, the dispersion relation curves of the lower and upper bands are symmetric to the point at the edge since the upper branch is simply the curve continuing the lower branch and folded back into the first Brillouin zone\footnote{Note that the symmetry would occur also if there was a band gap, e.g., for a superfluid in an optical lattice, then the symmetry point would be the one just in the middle of the band gap}. Therefore, in an interval around the pair of phonon modes $+q$ and $-q$, that is a continuum set of pairs $q \lesssim \pm k/2$ on the lower branch and $\mp k/2 \lesssim q$ on the upper branch fulfills both the momentum and energy conservation laws. This gives rise to an enhanced effective reservoir density of modes. The phonon energies at the edge are close to $\omega_R/4$, slightly raised due to collisions, therefore the peaks are expected at the values of the control parameter $y$ which lead to a polariton frequency at about $\frac{1}{2} \omega_R$. One can check by looking at the monotonous function shown in Fig.~\ref{fig:PolaritonSpectrum} that, below threshold, this occurs indeed at about $y/y_{\rm crit}\approx 0.8$.  Similar ``resonance'' of the soft mode frequency with phonons at the band edge occurs above threshold as is shown in the Figure. The other, smaller peak is of different origin, it arises form the overlap integrals at a certain shape of the condensate.

The reservoir density of modes is not infinite due to the finite phonon decay rate $\gamma_{1 \, q} + \gamma_{2 \, q} \equiv \epsilon$ blurring the sharpness of the energy conservation condition. Note that the precise shape of the damping rate as a function of $y$ slightly depends on the phenomenologically chosen value of  $\epsilon$, which reflects the role of this latter in the spectral density of reservoir modes. 

\begin{figure}[htb]
\begin{center}
\includegraphics[width=\columnwidth]{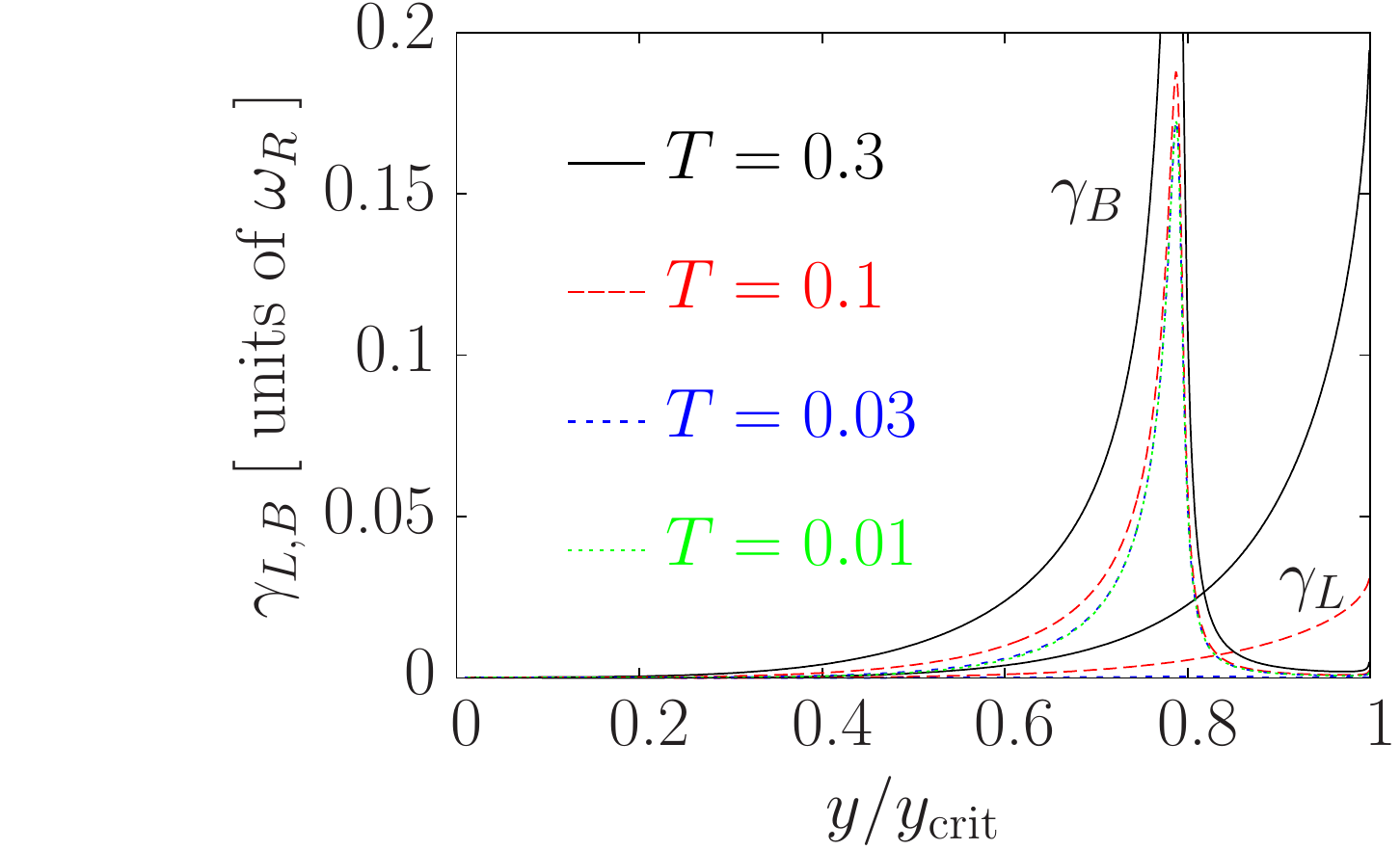}
\end{center}
\caption{(Color online) Dependence of the Landau- and Beliaev-damping rates on the temperature. For temperatures below the recoil frequency, the Landau damping does not suppress the Beliaev damping peak. Parameters are the same as for Fig.~\ref{fig:Beliaev_Fugges_kuszob_folott_abra}, and $\epsilon=0.01$.}
\label{fig:Landau_Beliaev_T_fugges_abra}
\end{figure}
The temperature dependence of the Landau- and Beliaev-damping rates is shown in Fig.~\ref{fig:Landau_Beliaev_T_fugges_abra}. The Landau-damping rate vanishes at zero temperature, but grows quickly as the temperature is increased. One can see that the Beliaev damping dominates in the whole range for temperatures up to $T=0.1\omega_R$, and the peak is significant even for higher temperatures $0.3 \omega_R \lesssim T$. 

\begin{figure}[ht]
\begin{center}
\includegraphics[width=\columnwidth]{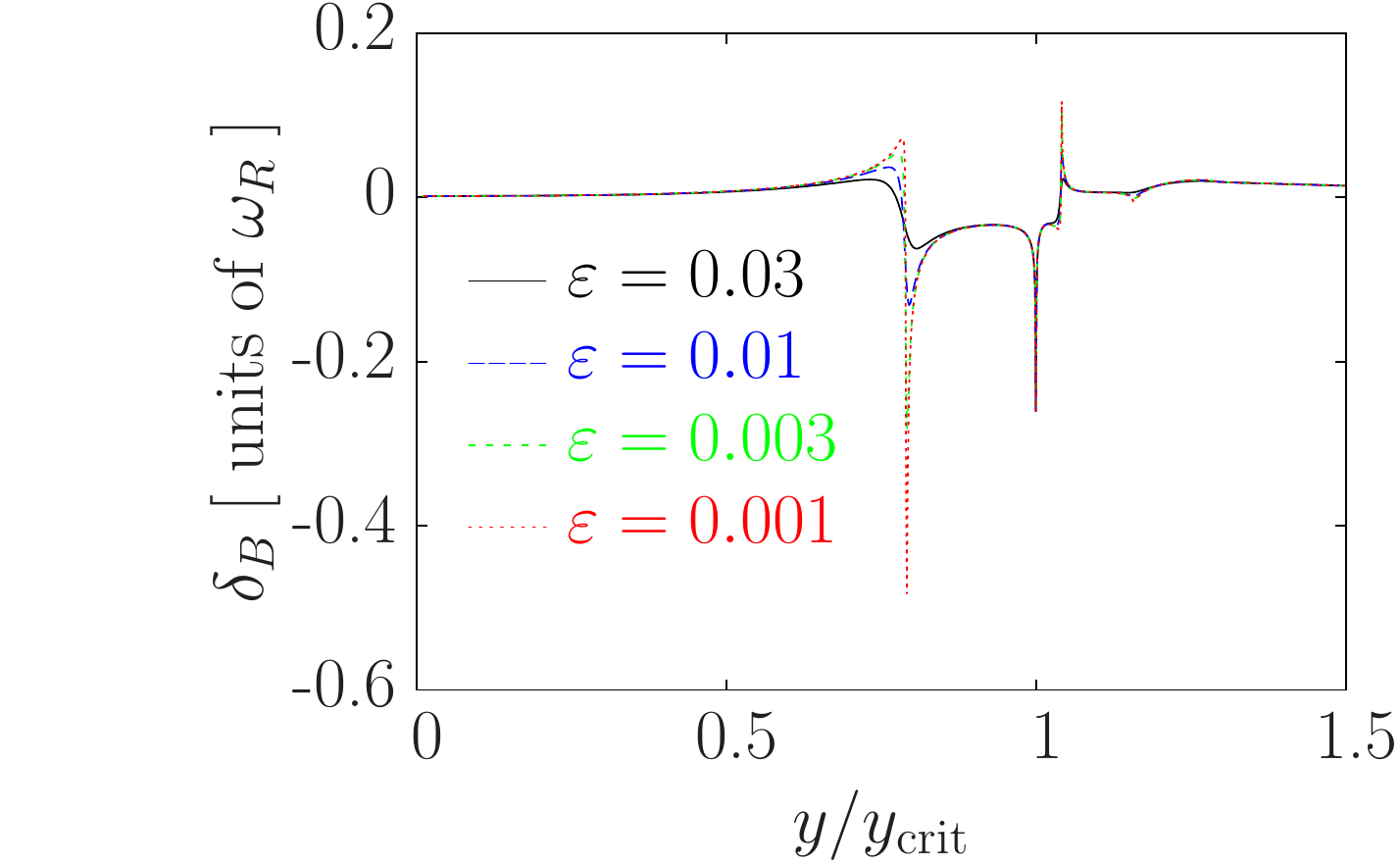}
\end{center}
\caption{(Color online) Beliaev frequency shift as a function of the normalized control parameter, indicating significant modification of the bare polariton frequency in conjunction with the enhanced damping rate. Parameters are the same as for Fig.~\ref{fig:Beliaev_Fugges_kuszob_folott_abra}.}
\label{fig:Beliaev_frequency_shift_abra}
\end{figure}

Let us also evaluate the real part of the self-energy in Born-Markov approximation, which is shown in Fig.~\ref{fig:Beliaev_frequency_shift_abra}.
The frequency shift is thus significant in the vicinity of the damping rate maximum. This result reveals that evaluating the self-energy at the bare polariton frequency may be very approximative. Therefore, in a next step, instead of the the Born approximation of the poles in Eq.~(\ref{eq:BornMarkovPole}), we consider the poles of the Green's functions arising from the zeros of the denominator  in Eq.~(\ref{eq:PolaritonGreen}).  

Before proceeding along this line, it is noteworthy to consider the dependence of the damping rate and frequency shift on the superfluid density $N_c/L$. Because of the summation in Eqs.~(\ref{eq:LandauDamping}) and (\ref{eq:BeliaevDamping}), there is an apparent factor of  the inverse of the density $\frac{1}{N_c} \sum_q$, however, the summands involve the square of the coupling constants $g_q^{L}$ or $g_q^{B}$ which are proportional to the density. Altogether the frequency shift and decay rates scale linearly with the superfluid density.

\subsection{Strong polariton-phonon coupling}

In order to get around the limitation of the Born-Markov approximation, namely that it assumes that the pole of the polariton Green's function is only shifted by a small amount due to the interaction with the phonons, which proved to be too strict, we look for the analytic structure of the Green's function directly and search the locations of the exact poles. We restrict ourselves only for Beliaev damping as this is the relevant damping channel at low temperatures. This way the analysis becomes easier and the interplay between the polariton and phonons is more transparent.

\begin{figure}[tb!]
\begin{center}
\includegraphics{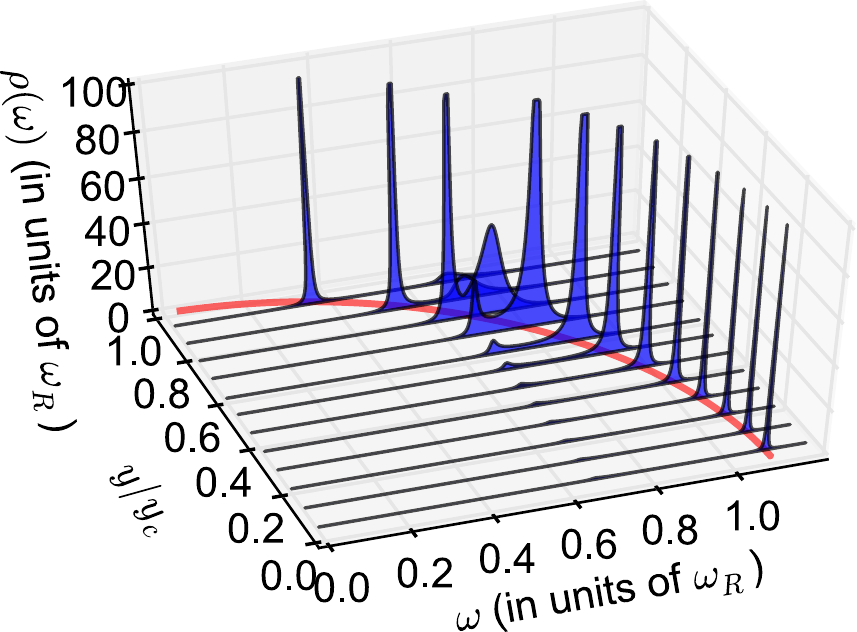}
\end{center}
\caption{(Color online) Spectral function of the polariton mode for various driving strengths $y$. The vertical range is truncated thus the high peaks, which are the dominantly polariton ones, are cut. These peaks lie quite precisely on the thick line drawn in the bottom plane, which is the polariton eigenfrequency in the Bogoliubov approximation, c.f. the curve in Fig.~\ref{fig:PolaritonSpectrum}, except for the range around $y/y_{\rm crit} \sim 0.8$. As the increasing control parameter approaches this range,  another peak grows up, which indicates that a  significant phonon component mixes to the polariton, and an avoided crossing can be observed.}
\label{fig:GreenFunctions}
\end{figure}
First let us define the spectral function for real frequencies, $\rho(\omega)\equiv-2\mathrm{Im }G(\omega)$ from which the retarded Green's function can be obtained in the usual way.
\begin{equation}
\label{eq:grspectrep}
G(\omega)=\lim_{\eta=0^+}\int_{-\infty}^\infty\frac{d\omega'}{2\pi}\frac{\rho(\omega')}{\omega-\omega'+i\eta}
\end{equation}
 A peak in the spectral function implies an elementary excitation, whose energy corresponds to the location of the peak, and its inverse lifetime to the width of the peak. 
One can directly evaluate the Green's function for real frequencies by using the same method for the evaluation of the sums in the self-energy functions (\ref{eq:Selfenergy}) as that we adopted for the Born-Markov approximation. Figure \ref{fig:GreenFunctions} presents the spectral function for various values of the control parameter.  It is clear that there are two significant peaks and an avoided crossing when the control parameter ($y/y_{\rm crit}$) is scanned between 0 and 1. At the extremes of the control parameter, one of the peaks can be attributed to the polariton mode, the other to the phonon bath.  The avoided crossing unambiguously signifies that a strong coupling between the polariton mode and the ensemble of phonon modes takes place. In other words,  the dynamics cannot be interpreted simply as a single dressed oscillator mode. It is strikingly unexpected that the polariton and  the phonons have such a considerable effect on each other.

The spectral function has a finite support in $\omega$, as the Beliaev self-energy \eqref{eq:Selfenergy} is integrated for the first Brillouin zone, where the real part of   $\omega^B_q$ is bounded. At the edges of the support the spectral function exhibits a peak. This peak can be attributed to the phonons and it is quite asymmetric, it has a sharp edge and a smooth fall-off. The other peak, corresponding to the polariton mode is of Lorentzian shape.   

\begin{figure}[tb!]
\begin{center}
\includegraphics{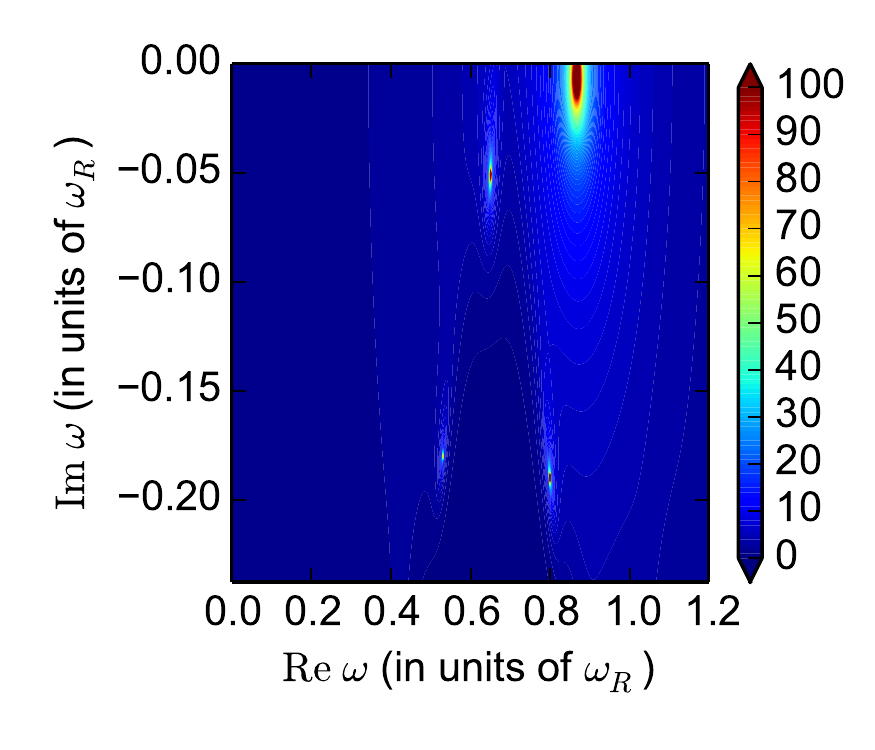}
\end{center}
\caption{((Color online) Complex analytic continuation of the retarded Green's function of the polariton mode for driving strength $y=0.629y_{\rm crit}$. The largest peak close to the real axis is dominantly the polariton mode. The phonon bath is represented by the multiple smaller peaks. The one closest to the large peak yields a strong phonon-polariton coupling influencing significantly the dependence of the polariton frequency as a function of the control parameter $y$ shown in Fig.~\ref{fig:GreenFunctions}.}
\label{fig:Green_z}
\end{figure}
To determine the position and the width of the peaks of the spectral function one analytically continues the retarded Green's function to the lower half of the complex plane $\omega \rightarrow z$ with ${\rm Im}\{ z \} < 0$. Poles encountered close to the real axis correspond to the excitations.  We carried out the analytic continuation  by solving the Cauchy-Riemann equations and propagating the solution gradually downward from the real axis.

\begin{figure}[tb!]
\begin{center}
\includegraphics[width=\columnwidth]{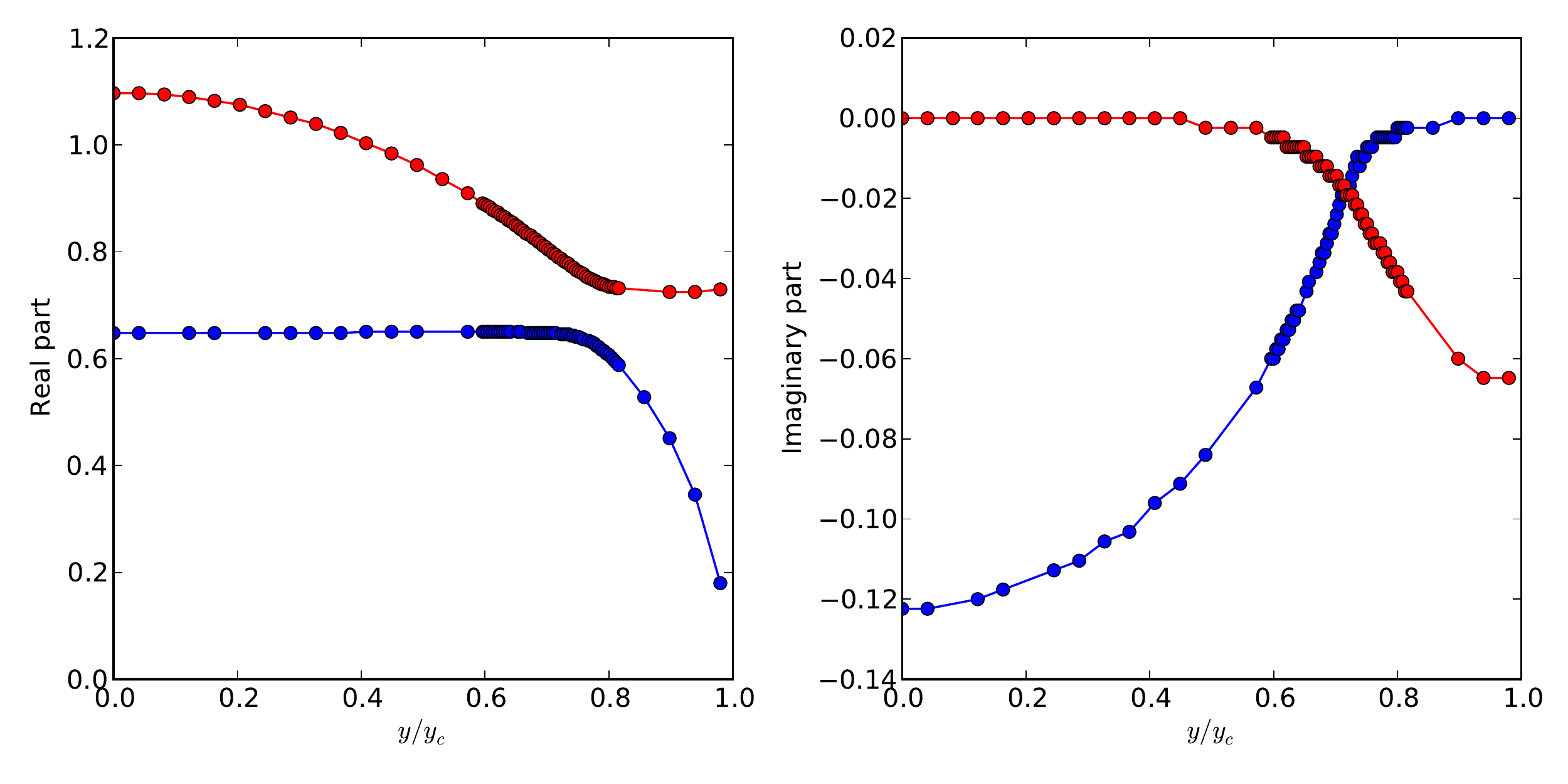}
\end{center}
\caption{(Color online)The real (left) and imaginary (right) parts of the two most relevant poles of the Green's function. Well-resolved avoided crossing can be seen in the real part, indicating a considerable mixing of the polariton with a collective phonon mode. The smaller imaginary part can be associated with the polariton damping rate which is then reduced compared with the Born-Markov prediction. }
\label{fig:PolesPosition}
\end{figure}

Since the spectral function has only finite support with a sharp fall-off, its endpoints correspond to branch points in the analytic continuation of the retarded Green's function. Therefore there is no unique analytic continuation to the whole complex plane. One can insert a single branch cut parallel to the real axis and between the branch points, or alternatively, take the function analytic between the branch points and insert two cuts connecting each branch point with the point infinitely far away.
To avoid such difficulties, we assume phonon modes at all real frequencies coupled extremely weakly to the polariton, thereby extending the finite cut along the whole straight line parallel to the real axis. Technically it means that we smooth out the spectral function a bit around the branch points. Then the analytic continuation is unambiguous on the lower half plane. We numerically computed the analytic function $\Sigma^{B}(z)$ and the corollary retarded Green's function $G(z)$ which,  for illustration purposes, is shown in Fig.~\ref{fig:Green_z} for a selected value of the control parameter $y$. The two-dimensional plot shows the pole corresponding to the polariton soft mode, and also other poles originating from the phonon bath. Since the spectral density of phonons is not a Lorentzian, there appears several poles of which the one closest to the real axis is the most relevant.
This is plotted in Fig.~\ref{fig:PolesPosition} which can then be considered a generalization of the result in Figs.~\ref{fig:Beliaev_frequency_shift_abra} and  \ref{fig:Beliaev_Fugges_kuszob_folott_abra}.
The real part manifests the avoided crossing, with a resolution much larger than the one used in Fig.~\ref{fig:GreenFunctions}, which demonstrates the strong coupling between the polariton and the phonons. The imaginary part reveals that the coupling to the polariton mode leads to a considerable narrowing of the effective width of the phonon bath. This effect is obviously beyond the usual Markov approximation assuming an inert reservoir. The smaller decay rate can be associated with the polariton. There is a peak at the crossing, however, the rate itself is an order of magnitude smaller than the one obtained by the Born-Markov approximation in the previous subsection.

\section{Summary}
\label{sec:Conclusions}

We studied a composite system which consists of a laser-driven Bose-Einstein condensate and a single-mode optical resonator. First, we determined the elementary excitations of this system using a Bogoliubov-type mean field analysis, which is given by the linear part of the basic equations \eqref{v_w_eq_of_motion} of the theory presented in this paper. We found that the atomic annihilation operators with zero quasi-momentum hybridize with the annihilation operator of the cavity field and after the diagonalization of $F_{\mu \nu}$ these lead to polariton excitations. The atomic annihilation operators with nonzero quasi-momentum do not couple to the photons at linear order, so they lead,  after the diagonalization of $G_{\mu \nu} (q)$, to the usual phonon excitations of the condensate. Since we are interested in the nonlinear polariton-phonon interaction, we have to go beyond the usual Bogoliubov approximation and consider the effects of the nonlinear terms in \eqref{v_w_eq_of_motion}.  From these terms, we constructed an effective Hamiltonian, c.f.~Eq.~\eqref{full_effective_H}, which contains the polariton and phonon operators as the basic constituents and describes their interaction. The effective Hamiltonian was simplified by restricting the system to the soft mode of the self-organization phase transition and to the relevant phonon bands. The simplified Hamiltonian contains two different types of interaction, called Landau and Beliaev processes, which are visualized in Fig.~\ref{fig:PhononSpectrum}.

Since the effective Hamiltonian contains third order terms, it is not possible to solve the problem exactly. If we try to solve it using the equation of motion of the polariton  Green's function, then we run into an infinite hierarchy of equations: three point functions appear in the equations of two point functions and so on. To deal with this problem, we use a bosonization approximation which relies on that the phonons are close to be in a thermalized state. By rendering the effective Hamiltonian bilinear in the variables, the equations of the Green's functions can be solved straightforwardly. As a result, we obtain the self-energies in Eq.~\eqref{eq:Selfenergy}.

From the self-energy, we can take two distinct paths to evaluate the damping rate of the polaritons due to the phonon bath. As the simplest one, we can use the Born-Markov approximation where the self-energy is evaluated at the bare frequency of the polariton. The imaginary part gives the required damping rate, whereas the real part corresponds to a frequency shift. This latter turned out to be significant with respect to the bare frequency.  This motivated us for using another, more accurate approach. The frequency dependence of the self-energy on the real frequency axis has been retained and we performed numerically an analytic continuation to the lower half plane. We found the location of the pole, interestingly, however, there were two relevant poles. One of them corresponds to the expected polariton soft mode, the other one to a collective mode within the phonon bath. As the strength of the laser pump is varied, there is an avoided crossing between these two poles, which indicates a significant back action of the polariton to the phonon bath. 

\section*{Acknowledgements}

This work was supported by the Hungarian National Office for Research and Technology under the contract ERC\_HU\_09 OPTOMECH, the Hungarian Academy of Sciences (Lend\"ulet Program, LP2011-016), and the Hungarian Scientific Research Fund (grant no. PD104652). G.Sz. also acknowledges support from the J\'anos Bolyai Scholarship.

\appendix

\section{Connection between $V^{\beta \, \gamma}_{\alpha}$ and $W^{\beta \, \gamma}_{\alpha}$}
\label{V_W_connection_appendix}

In this appendix, we derive an equation which connects $V^{\beta \, \gamma}_{\alpha}$ and $W^{\beta \, \gamma}_{\alpha}$. This equation is needed to prove equation \eqref{ABCD_MNO}.

The commutation relations of $\tilde{v}_{\alpha}$ and $\tilde{w}_{\beta}(q)$ are given by the following formulas:
\begin{subequations}
\begin{align}
\comm{\tilde{v}_{\alpha}}{\tilde{w}_{\beta}(q)} &=0 \\
\comm{\tilde{v}_{\alpha}}{\tilde{v}_{\beta}} &= \left( \Omega \cdot \Gamma \right)_{\alpha \, \beta} \\
\comm{\tilde{w}_{\alpha}(-q)}{\tilde{w}_{\beta}(q)} &= \left( \Omega \cdot \Gamma \right)_{\alpha \, \beta} \\
\comm{\tilde{w}^{\dag}_{\alpha}(q)}{\tilde{w}_{\beta}(q)} &= -\Omega_{\alpha \, \beta}
\end{align}
\end{subequations}
These formulas should hold for all time $t$. Now, if we take the time derivative of the first commutator listed here, we can deduce the relationship between $V$ and $W$:
\begin{multline} \label{V_W_connection}
-\sum_{\alpha} \, V_{\mu}^{\alpha \, \beta} \cdot \Omega_{\alpha \, \nu} + \sum_{\alpha \, \delta} \, V_{\mu}^{\alpha \, \delta} \cdot \Gamma_{\alpha \, \beta} \left( \Omega \, \Gamma \right)_{\delta \, \nu}  \\
+ \sum_{\alpha} \, W_{\nu}^{\alpha \, \beta} \cdot \left( \Omega \cdot \Gamma \right)_{\mu \, \alpha} =0
\end{multline} 

To prove \eqref{ABCD_MNO}, we also need the following formulas, which come from the application of the symmetry $\Gamma$:
\begin{subequations}
\begin{align}
V_{\mu'}^{\alpha \, \beta}  &= - \sum_{\mu} \, \Gamma_{\mu' \, \mu} \cdot V^{\beta \, \alpha \, ^*}_{\mu}                    \\
V_{\mu}^{\beta' \, \alpha'} &= + \sum_{\alpha' \, \beta'} V_{\mu}^{\alpha \, \beta} \cdot \Gamma_{\alpha \, \alpha'} \cdot \Gamma_{\beta \, \beta'}  \\
W_{\mu'}^{\alpha' \, \beta'} &= - \sum_{\mu \, \alpha \, \beta} \, \Gamma_{\mu' \, \mu} \cdot W^{\alpha \, \beta \, ^*}_{\mu}
\cdot \Gamma_{\alpha \, \alpha'} \cdot \Gamma_{\beta \, \beta'}
\end{align}
\end{subequations}

\bibliography{longbeliaev}

\end{document}